\documentclass[12pt]{article}
\usepackage{graphicx}
\usepackage{epstopdf}

\usepackage[T1]{fontenc}
\usepackage{amsmath}
\usepackage{amssymb}
\usepackage{times}
\usepackage{txfonts}
\usepackage[mathlines,displaymath]{lineno}
\usepackage{pdflscape}
\usepackage{afterpage}
\usepackage{rotating}
\DeclareMathAlphabet{\mathbold}{OML}{txr}{b}{it}

\usepackage{epstopdf}
\usepackage{epsfig}
\usepackage{hhline}
\usepackage{cite}
\usepackage{subfig} 
\usepackage{multirow}
\RequirePackage{filecontents}

\newlength{\dinwidth}
\newlength{\dinmargin}
\newcommand{\pom}{{I\!\!P}}

\newcommand{\qsq}{\ensuremath{Q^2} }
\newcommand{\mxq}{\ensuremath{M_{X}^{2}} }

\newcommand{\xpom}{x_{\pom}}
\newcommand{\zpom}{z_{\pom}}

\newcommand{\asmz}{\ensuremath{\alpha_s(M_Z)}}

\begin{document}

\begin{titlepage}

\noindent
\begin{flushleft}
{\tt DESY 14-200    \hfill    ISSN 0418-9833} \\
{\tt November 2014}                  \\
\vspace{\baselineskip}
\end{flushleft}

\vspace{2cm}
\begin{center}
\begin{Large}

{\bf Measurement of Dijet Production in Diffractive Deep-Inelastic $\mathbold{ep}$ Scattering at HERA}

\vspace{2cm}

H1 Collaboration

\end{Large}
\end{center}

\vspace{2cm}

\begin{abstract}
A measurement is presented of single- and double-differential dijet cross sections
in diffractive deep-inelastic $ep$ scattering at HERA 
using data collected by the H1 experiment corresponding to an integrated luminosity
of $290\ \rm{pb^{-1}}$.
The investigated phase space is spanned by the photon virtuality
in the range of $4<Q^{2}<100\ \rm GeV^{2}$ and by the fractional proton
longitudinal momentum loss $\xpom<0.03$.
The resulting cross sections are compared with next-to-leading order QCD predictions
based on diffractive parton distribution functions and 
the value of the strong coupling constant is extracted.
\end{abstract}

\vspace{1.5cm}

\begin{center}
Submitted to \it{JHEP} 
\end{center}

\end{titlepage}

 

V.~Andreev$^{21}$,             
A.~Baghdasaryan$^{33}$,        
K.~Begzsuren$^{30}$,           
A.~Belousov$^{21}$,            
V.~Boudry$^{24}$,              
G.~Brandt$^{45}$,              
V.~Brisson$^{23}$,             
D.~Britzger$^{10}$,            
A.~Buniatyan$^{2}$,            
A.~Bylinkin$^{20,42}$
% ...


\bigskip{\it
% ...
$ ^{2}$ School of Physics and Astronomy, University of Birmingham,
          Birmingham, UK$^{ b}$ \\
% ...
$ ^{10}$ DESY, Hamburg, Germany \\
% ...
$ ^{20}$ Institute for Theoretical and Experimental Physics,
          Moscow, Russia$^{ i}$ \\
 $ ^{21}$ Lebedev Physical Institute, Moscow, Russia \\
% ...
$ ^{23}$ LAL, Universit\'e Paris-Sud, CNRS/IN2P3, Orsay, France \\
 $ ^{24}$ LLR, Ecole Polytechnique, CNRS/IN2P3, Palaiseau, France \\
% ...
$ ^{30}$ Institute of Physics and Technology of the Mongolian
          Academy of Sciences, Ulaanbaatar, Mongolia \\
% ...
$ ^{33}$ Yerevan Physics Institute, Yerevan, Armenia \\
% ...
$ ^{42}$ Also at Moscow Institute of Physics and Technology, Moscow, Russia \\
% ...
$ ^{45}$ Department of Physics, Oxford University,
          Oxford, UK$^{ b}$ \\

\bigskip
% ...
$ ^b$ Supported by the UK Science and Technology Facilities Council,
      and formerly by the UK Particle Physics and
      Astronomy Research Council \\
% ...
$ ^i$ Russian Foundation for Basic Research (RFBR), grant no 1329.2008.2
      and Rosatom \\
% ...
}

\section{Introduction}

In deep-inelastic scattering (DIS),
diffractive reactions of the type $ep\to eXY$, where $X$ is a high-mass hadronic final state and
$Y$ is either the elastically scattered proton or its low-mass excitation,
represent about $10\%$ of the events at HERA and provide rich experimental input for testing
quantum chromodynamics (QCD) in the diffractive regime. These processes can be understood 
as probing by a virtual photon
emitted from the beam lepton a net colour singlet carrying vacuum quantum numbers (a pomeron)~\cite{pomeranchuk,goulianos}.
Due to the colourless exchange the systems $X$ and $Y$ are separated by a rapidity
interval free of hadronic activities.
In these processes at least one hard scale is involved such that perturbative QCD (pQCD) can be applied.

According to the QCD collinear factorisation theorem~\cite{trentadue},  
calculations of diffractive cross sections factorise into process dependent hard scattering coefficient functions
and a set of process independent diffractive parton distribution functions (DPDFs).  
While the hard scattering coefficient functions are calculable in pQCD,
the DPDFs have to be determined from QCD fits to the measured inclusive diffractive cross sections.
In such QCD fits~\cite{fitb}, DGLAP evolution~\cite{gribov,dokschitzer,altarelli} of the DPDFs is assumed.
The QCD factorisation theorem is proven to hold for inclusive and dijet diffractive processes~\cite{collins},
assuming high enough photon virtuality such that higher twist effects can be neglected.
The DPDFs are experimentally determined by assuming an additional
factorisation of the DPDFs dependence on the scattered proton momentum from the dependence on the other variables, ascribed to the structure of the colourless exchange.
This assumption is known as proton vertex factorisation.
A pomeron flux in the proton is introduced and universal parton densities
are attributed to the diffractively exchanged object.
Many measurements of diffraction in DIS suggest the validity 
of the proton vertex factorisation assumption in DIS~\cite{rfact_1,rfact_2,rfact_3,fitb}.

In leading order the inclusive diffractive cross section in $ep$ scattering
is proportional to the charge-squared weighted sum of the quark distribution functions in the pomeron,
while 
its gluon content can be determined only indirectly via scaling violations.
As events with two jets (dijets) are readily produced in gluon-induced processes,
measurements of diffractive dijet cross sections are 
sensitive to the value of the strong coupling $\alpha_s$  and
to the gluon content of the pomeron.
The production of dijets in diffractive DIS has previously been studied
at HERA using either the large rapidity gap (LRG) method~\cite{jetdiff_1,jetdiff_3,jetdiff_zeus} or
by direct detection of the outgoing proton~\cite{jetdiff_2}.

In this paper cross section measurements of dijet production in diffractive $ep$ scattering
are presented, based on data collected in the years 2005-2007 with the H1 detector at HERA.
Diffractive events are selected
by means of the LRG method, requiring a clear separation in rapidity of the final state systems
$X$ and $Y$. 
The measured cross sections are compared to next-to-leading order (NLO) QCD predictions evaluated with input DPDFs
determined in previous inclusive diffractive measurements by the H1 collaboration~\cite{fitb}.

The present analysis is based on the full HERA-II data sample resulting in significantly increased
statistics with respect to previous analyses. Furthermore, the cross sections are determined
using a regularised unfolding procedure which fully accounts for efficiencies, migrations and  correlations among the measurements. 
The measured dijet cross sections are used to extract the strong coupling constant $\alpha_{s}$ 
in diffractive DIS processes for the first time.


\section{Kinematics}

\begin{figure}[ht]
\centering
\includegraphics[width=.5\textwidth]{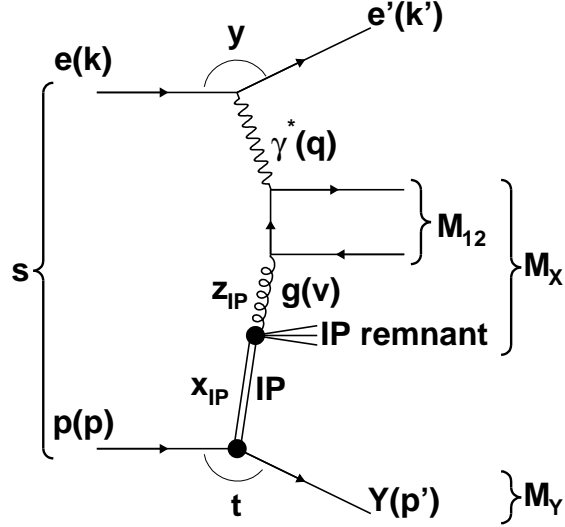}
\caption{Leading order diagram for the production of dijets in diffractive DIS.}
\label{fig:feyn}
\end{figure}

A leading order (LO) diagram of boson-gluon fusion, which is the dominant process for the production of two jets in diffractive DIS,
is depicted in Figure~\ref{fig:feyn}. 
The incoming electron\footnote{In this paper the term "electron" is used generically to refer to both electrons and positrons.} 
of four-momentum $k$ interacts with 
the incoming proton of four-momentum $p$ via the exchange
of a virtual photon of four-momentum $q=k-k'$. The outgoing proton or its low-mass dissociation state carries
four-momentum $p'$. 
The DIS kinematics is described
by the following set of variables:
\begin{equation}
\label{_kine_dis}
\qsq=-q^{2}=(k-k')^{2}, \qquad x=\frac{\qsq}{2p\cdot q}, \qquad y=\frac{p\cdot q}{p\cdot k}, 
\end{equation}
where $\qsq$, $x$ and $y$ denote the photon virtuality,
the Bjorken-$x$ variable and the inelasticity of the process, respectively.
Conservation laws stipulate the relation $\qsq=xys$, where $s$ stands for the $ep$ centre-of-mass energy squared.

The kinematics of the diffractive exchange is described in terms of the additional quantities
\begin{equation}
\label{_kine_diff}
\xpom=\frac{q\cdot(p-p')}{q\cdot p}, \qquad t=(p-p')^{2}
\end{equation}
with $\xpom$ and $t$ being the longitudinal momentum fraction of the incoming proton carried by the pomeron and 
the squared four-momentum transfer at the proton vertex, respectively. 
The fractional longitudinal momentum of the pomeron transferred to the dijet system is given by
\begin{equation}
\label{_kine_jet}
\zpom=\frac{q\cdot v}{q\cdot(p-p')}=\frac{x}{\xpom},
\end{equation}
where $v$ is the four-momentum of the parton entering the hard interaction. 

\section{Monte Carlo Models and Fixed Order QCD Calculations}
\label{sec:model}
The RAPGAP event generator~\cite{rapgap} allows for the simulation of processes $ep\to eXY$ 
including both leading (pomeron) and sub-leading (reggeon) exchanges.
Assuming the proton vertex factorisation, the parton densities obtained
in the previous QCD analysis of inclusive diffractive data (H12006 Fit-B)~\cite{fitb}
are convoluted with leading order QCD matrix elements. 
Higher order QCD radiation effects are modelled via
initial and final state parton showers in the leading-$\log$ approximation~\cite{llog}.
Hadronisation is accounted for by making use of the Lund string model~\cite{lund} as implemented in PYTHIA~\cite{pythia}.

Within the diffractive selection based on the LRG method, the system $Y$ may also be a low mass dissociative system.
Proton dissociation events are simulated in the the range of $M_Y<20\ \mathrm{GeV}$ using the RAPGAP event generator, where  
$M_Y$ is the mass of the system $Y$.
Resonant contributions together with the continuum part of the $M_Y$ distribution are modelled similarly 
to the DIFFVM event generator~\cite{diffvm}. 
A small admixture of resolved $\gamma^{*}p$ scattering is included in fixed LO mode of jet production
in the low $\qsq$  region~\cite{resgamma}. The resolved photon contribution is simulated
with the RAPGAP event generator using the SaS-G PDF set~\cite{sasg2dfit} as the input PDF of the photon. 
QED radiation effects are simulated with the HERACLES~\cite{heracles} program interfaced to RAPGAP.
Besides the Born level contribution, the simulated cross sections
include contributions from initial and final state emission of real photons from the electron, from vertex corrections
as well as from self energy diagrams.
As the H12006 Fit-B DPDF set has previously been observed to underestimate the data in the low $\qsq$ region,
a weighting is applied for $\qsq<7\ \mathrm{GeV^{2}}$,
parametrised as the ratio of the data in~\cite{fitb} to the Monte Carlo expectation based on the H12006 Fit-B DPDF set.

Background arising from non-diffractive DIS processes
is also simulated with the RAPGAP event generator
using its inclusive mode together with the CTEQ6L PDF set~\cite{cteq}.

The MC simulation is used to correct the data for detector effects. 
The generated events undergo the full GEANT~\cite{geant} simulation of the H1 detector
and are analysed in the same way as the real data.
In order to describe the measured distributions, the diffractive MC
is reweighted in several variables as discussed in~\ref{subsec:data}. 

QCD predictions of the dijet cross sections at the parton level are evaluated at NLO
using the NLOJET++ program~\cite{nlojetprog,nlojet}. 
The NLO pQCD predictions are calculated in the $\overline{\rm MS}$-scheme with five active flavors.
The two-loop approximation of the renormalisation group equation is used for the running of the strong coupling constant
with a coupling strength of $\alpha_s(M_Z)=0.118$.
The cross sections are evaluated in intervals of $\xpom$, effectively
replacing the beam proton by a pomeron (slicing method).
The H12006 Fit-B DPDF set is used in the calculation. 
The renormalisation and factorisation scales $\mu_{r}$ and $\mu_{f}$ are provided by the photon virtuality and  
the average transverse momentum of the leading and sub-leading jet, $\langle p^{\ast}_{\rm T}\rangle$, 
in the $\gamma^{\ast}$-$p$ centre-of-mass frame and are defined as $\mu_{r}=\mu_{f}=\sqrt{\langle p^{\ast}_{\rm T}\rangle^{2}+\qsq}$.
The uncertainty on the prediction due to missing higher orders is estimated by simultaneous variation of the renormalisation
and factorisation scales by factors of $0.5$ or $2$. 
An uncertainty on the NLO prediction from the experimental uncertainties on
the DPDF set is obtained using the eigenvector decomposition
of the uncertainties of the H12006 Fit-B DPDF set.
This uncertainty is propagated to the NLO prediction
using the sign-improved formulae for error propagation~\cite{dpdferr}.
A significant contribution to the uncertainty of the H12006 Fit-B set
originates from the restriction of the input data to $\zpom<0.8$ and the extrapolation of the DPDF to $\zpom>0.8$.

Whereas the measured cross sections are compared to the predictions obtained by the slicing method,
an alternative method of adapting the NLO calculations
for diffractive DIS is used in the $\alpha_s$ extraction.
In order to provide theory predictions  
with different values of $\asmz$, the fastNLO method~\cite{H1fastNLO,fastNLO,fnlo} is used.
Cross section predictions are obtained by folding tabulated matrix
elements obtained from NLOJET++~\cite{nlojetprog,nlojet} with the DPDF
parametrisation.
The matrix elements are determined as a function of the observable of
interest, the factorisation scale $\mu_F$ and the convolution variable $x$.
The relation $x=\xpom\zpom$ is used when folding with the DPDF.
This way predictions can be obtained for different choices of DPDFs, of $\alpha_s$ and of
the renormalisation and the factorisation scales
without having to calculate the matrix elements all over again.
Settings identical to the slicing method are used for parameters such
as renormalisation and factorisation scales or DPDF set and 
very good numerical agreement with the slicing method is found.
The uncertainty on the prediction due to missing higher orders
is estimated by varying the scales by a factor $f$, where $0.5<f<2$.

Since the measured cross sections are given at the level of stable hadrons, the QCD predicted cross sections
have to be corrected for effects of initial and final state parton showers, hadronisation and fragmentation.
These corrections are determined for each of the measured cross sections as the ratio of hadron to parton level cross sections,
predicted with the RAPGAP event generator. 
Two distinct models of parton showers, 
the leading-$\log$ approximation and the colour dipole model as implemented in the ARIADNE program~\cite{ariadne},
are used in this calculation. 
In each measurement interval the resulting correction is
taken as the average of the values predicted by the two
models and the uncertainties on the correction factors
are taken as half the difference of the two predictions.
The hadron level cross sections are on average about $5\%$  higher than the parton level cross sections.
The total uncertainty on the NLO QCD predictions is obtained as the quadratic sum of the uncertainties
from scale variation, DPDF fit and hadronisation uncertainties.

\section{Experimental Technique}
\subsection{H1 detector}

A detailed description of the detector can be found elsewhere~\cite{det_1}. 
Here only those detector components 
relevant for the present analysis are briefly described.  
A right-handed coordinate system with the origin at the nominal interaction point
and with the $z$-axis pointing in the proton beam direction is conventionally chosen
as the laboratory frame. The polar angle $\theta$ is measured with respect to the $z$-axis,
while the direction in the $x$-$y$ plane is defined by the azimuthal angle $\phi$.
The pseudorapidity is defined as $\eta=-\ln\tan(\theta/2)$. 

The liquid argon (LAr) sampling calorimeter~\cite{lar} is located inside a $1.15\ \rm T$ solenoidal field 
and covers the polar angular range $4^\circ\!<\!\theta\!<\!154^\circ$.
The energy resolutions for electromagnetic and hadronic showers as determined in test beam measurements~\cite{testbeam_1,testbeam_2}
are $\sigma(E)/E \propto 11\%/\sqrt{E/\rm GeV}\oplus 1\%$ and
$\sigma(E)/E \propto 50\%/\sqrt{E/\rm GeV}\oplus 2\%$, respectively.
The energy and scattering angle of the scattered electron is measured in a scintillating fibre calorimeter SpaCal~\cite{spac1,spac2}
with a resolution of $\sigma(E)/E \propto 7\%/\sqrt{E/\rm GeV}\oplus 1\%$.
The precision of the energy scale is $1\%$ covering the polar angular range $154^\circ\!<\theta_{e'}<\!174^\circ$. 
The measurement of the polar angle of the scattered electron $\theta_{e'}$ is improved by means of a backward proportional chamber (BPC).
The precision of the polar angle measurement is $1\ \rm mrad$.

Trajectories of charged particles are measured with the central tracking detector (CTD) 
located inside the LAr calorimeter 
with a transverse momentum resolution of 
$\sigma_{p_T}/p_T\simeq 0.2\ \%\cdot\ p_{T}/\rm GeV\oplus1.5\%$
in the polar angular range of $15^\circ\!<\!\theta\!<\!165^\circ$.

The information from CTD and LAr is used for the reconstruction
of the system $X$.
The interaction vertex position is determined event-by-event 
using the particle trajectories measured in CTD.

The following H1 forward detectors are used in the LRG selection of diffractive events.
The forward muon detector (FMD) consists of six proportional chambers which are grouped into
two three-layer sections separated by a toroidal magnet. Although the nominal
coverage of FMD is $1.9<\eta<3.7$, particles with pseudorapidity up to $\eta\! \sim\! 6.5$
can be detected indirectly through their interactions with the beam transport system and detector support structures. 
The lead-scintillator Plug calorimeter is located at $z=4.9\ \rm m$ and covers the range $3.5<\eta<5.5$.
The very forward region is covered by the forward tagging system (FTS) comprising
scintillators surrounding the beam pipe. Only one station of FTS, situated at $z=28\ \rm m$
and covering the range $6.0<\eta<7.5$, is included in the present analysis. 

The instantaneous luminosity is monitored based on the rate of the Bethe-Heitler process $ep\to ep\gamma$.
The final state photon is detected by a photon detector located close to the beam pipe at $z=-103\ \rm m$. 
The precision of the integrated luminosity measurement is improved in a dedicated analysis of the QED Compton
process~\cite{qedclumi}.

\subsection{Reconstruction of observables}

The DIS observables $\qsq$, $x$ and $y$ are reconstructed using the electron-$\Sigma$ method ~\cite{esigma}.
Within this method, the photon virtuality $\qsq$ is reconstructed based on the measured four-momentum of the scattered electron,
while the inelasticity $y$ and Bjorken-$x$ are determined making use
of combined information from the  
hadronic final state (HFS) and the scattered electron.  

The four-momenta of the particles attributed to HFS 
are reconstructed using an algorithm which combines information
provided by the tracking system and the LAr calorimeter
by avoiding double counting of hadronic energy~\cite{peez:03,hellwig:04}. 
The calibration of the HFS energy scale derived in~\cite{kogler} is applied.
The performance of the calibration was studied by comparing 
the transverse momentum balance in data and MC in the kinematic domain of this analysis.
  
Jets are reconstructed in the $\gamma^{\ast}$-$p$ centre-of-mass frame
using the inclusive 
$k_{\rm T}$ jet algorithm ~\cite{kt} with the $p_{\rm T}$ recombination scheme as implemented in the FastJet program ~\cite{fastjet}. 
The jet distance parameter is set to $R=1.0$.
The transverse momenta and pseudorapidities of the leading
and sub-leading jets are denoted as $p^{\ast}_{\rm T,1}$, $\eta^{\ast}_{1}$ and $p^{\ast}_{\rm T,2}$, $\eta^{\ast}_{2}$, respectively\footnote{Observables in the $\gamma^{\ast}$-$p$ centre-of-mass frame are labelled with an asterisk.}.  
The invariant mass of the final state system $X$ is reconstructed as:
\begin{equation}
M_{X}=c(\eta_{max})\sqrt{P_{X}^{2}}, 
\end{equation}
where $P_{X}$ is the four-momentum of the system $X$ obtained as a vector sum of all particles contained 
in the HFS.
The MC simulation is used in order to derive the average correction for detector losses $c(\eta_{max})$, 
where $\eta_{max}$ is the pseudorapidity of
the most forward energy deposition above $800\ \rm MeV$ in the LAr calorimeter.
The momentum fractions $\xpom$ and $\zpom$ are reconstructed as: 
\begin{equation}
\label{_reco_xpom}
\xpom=\frac{\qsq + \mxq}{ys}
\end{equation}
and
\begin{equation}
\label{_reco_zpom}
\zpom=\frac{\qsq + M_{12}^{2}}{\qsq + \mxq},
\end{equation}
where $M_{12}$ is the invariant mass of the dijet system.  


Cross sections for dijet production in diffractive DIS are measured differentially with respect to the variables $\qsq$, $y$, $\xpom$, $\zpom$,
$p^{\ast}_{\rm T,1}$, $p^{\ast}_{\rm T,2}$, $\langle p^{\ast}_{\rm T}\rangle=(p^{*}_{\rm T,1}+p^{\ast}_{\rm T,2})/2$ and 
$\Delta\eta^{\ast}=|\eta^{\ast}_{1}-\eta^{\ast}_{2}|$. 

\subsection{Event selection}

The measurement is based on the H1 data collected
in the years $2005$ to $2007$ with a total integrated luminosity of $290\ \rm pb^{-1}$. 
The nominal beam energies of the protons and electrons are $E_{p}=920\ \rm GeV$ and $E_{e}=27.6\ \rm GeV$, respectively.

The longitudinal position of the reconstructed event vertex 
is restricted to the range $-35<z_{vtx}<35\ \rm cm$. 
DIS events are selected by the identification of the scattered electron in the backward calorimeter SpaCal.
The isolated energy deposit of electromagnetic structure with the highest transverse momentum is identified as scattered electron and 
has to have a measured energy of at least $9.5\ \rm GeV$. 

Only events accepted by a trigger combining signals induced by the scattered electron in the SpaCal 
with minimum track information of  the CTD  are used in the analysis.
The trigger efficiency related to the CTD condition is found to be
$98\%$-$99\%$, depending on the detector configuration
and is reproduced by the MC simulation within $2\%$.
The trigger efficiency related to the SPACAL condition is better than $99\%$.

Residual non-DIS background is dominated by photoproduction processes, 
where a hadron is misidentified as the scattered electron, whereas the true
scattered electron escapes detection due to its small scattering angle.
This background is reduced to a negligible level by demanding
$35<\sum_{i}(E-p_{z})_{i}<75\ \rm GeV$, where the sum runs over all HFS particles and the scattered electron candidate.
Elastic QED Compton scattering $ep\to e\gamma p$ introduces another background contribution
which is suppressed by rejecting configurations with two back-to-back clusters in SpaCal.
  
Diffractive events are identified with the LRG method 
which requires an empty interval in rapidity between the systems $X$ and $Y$. 
The low-mass system $Y$ is produced at very large pseudorapidities
and escapes detection. The diffractive signature is thus defined by
the systems $X$ (in the main detector) and $Y$ (undetected).
The energy of any cluster in the forward region
of the LAr calorimeter is required to be below the noise level of $800\ \rm MeV$,
which is ensured by demanding $\eta_{max}<3.2$.
The variable $\eta_{max}$ corresponds to the LAr cluster above
the noise threshold which has the largest pseudorapidity.
Information provided by the forward detectors FMD, FTS and the Plug calorimeter
is used in order to extend the gap to rapidities beyond the LAr acceptance
and in order to suppress the proton dissociation contribution.
These detectors are required to show no signal above noise level~\cite{mythesis}.
At high momentum fractions $\xpom$,
the system $X$ tends to extend into the direction of the outgoing system $Y$
and the experimental separation of the systems $X$ and $Y$ is not possible.
The LRG selection method is thus applicable only in the region of $\xpom\lesssim0.03$.
The sample of DIS events satisfying the LRG criteria is dominated by the diffractive exchange, 
as the system $X$ is isolated
in the main part of the H1 detector, while the system $Y$ escapes undetected
down the beam pipe. The signal is dominated by proton-elastic processes, $ep\to eXp$, however,
a small fraction of proton dissociation events is also accepted by the LRG selection. 
The LRG requirements impose restrictions on the mass and scattering angle
of the hadronic system $Y$. These correspond approximately to the requirements
$M_{Y}<1.6\ \rm GeV$ and $|t|<1\ \rm GeV^{2}$. 
Migrations in these variables are modelled using MC simulations.

\begin{table}[h!]
\begin{center}
\begin{tabular}{|c|c|c|}
\hline
& \bf{Extended Analysis Phase Space} & \bf{Measurement Cross Section Phase Space} \\
\hline
\multirow{2}{*}{DIS} & $3 < Q^{2} < 100 \ \mathrm{GeV^{2}}$ & $4 < Q^{2} < 100 \ \mathrm{GeV^{2}}$ \\
& $y < 0.7$ & $0.1 < y < 0.7$ \\
\hline
\multirow{3}{*}{Diffraction} & $x_{\pom} < 0.04$ & $x_{\pom} < 0.03$ \\
& LRG requirements & $|t|<1 \ \mathrm{GeV^{2}}$ \\
& & $M_{Y} < 1.6 \ \mathrm{GeV} $ \\
\hline
\multirow{3}{*}{Dijets} & $p_{\rm T,1}^{*} > 3.0 \ \mathrm{GeV}$ & $p_{\rm T,1}^{*} > 5.5 \ \mathrm{GeV} $\\
& $p_{\rm T,2}^{*} > 3.0 \ \mathrm{GeV} $ & $p_{\rm T,2}^{*} > 4.0 \ \mathrm{GeV} $ \\
& $-2 < \eta^{\rm lab}_{1,2} < 2$ & $-1 < \eta^{\rm lab}_{1,2} < 2$ \\
\hline
\end{tabular}
\end{center}
\caption{Summary of the extended analysis phase space and the phase space for the dijet cross sections measurements.}
\label{tab:ps}
\end{table}

Events are selected in a phase space which is extended compared to the measurement phase space
in order to improve the precision of the measurement 
by accounting for migrations at the phase space boundaries.
Events within the DIS phase space of $y<0.7$ and $3<\qsq<100\ \rm GeV^2$ are selected.
The events are required to have at least two jets in the pseudorapidity range $-2<\eta^{\rm lab}_{1,2}<2$
and transverse momenta greater than $3\ \rm GeV$ in the $\gamma^{\ast}$-$p$ centre-of-mass frame.

The measurement phase is defined by the DIS requirements of $0.1<y<0.7$ and $4<\qsq<100\ \rm GeV^2$.
The pseudorapidity of jets is restricted in the laboratory frame to $-1<\eta^{\rm lab}_{1,2}<2$ to
ensure the jets to be contained well within the central detector. 
The transverse momenta
of the leading and sub-leading jets are required to be larger than $5.5\ \rm GeV$ and $4.0\ \rm GeV$, respectively.
The extended phase space and the measurement phase space definitions are summarised in table~\ref{tab:ps}.
The total number of events accepted by the LRG selection criteria together with the DIS and jet requirements is $\sim\! 50000$
and $\sim\! 15000$ for the extended and measurement phase space, respectively.

\subsection{Corrections to the data}
\label{subsec:data}

\begin{figure}[ht]
\centering
\subfloat{
\includegraphics[width=.48\textwidth]{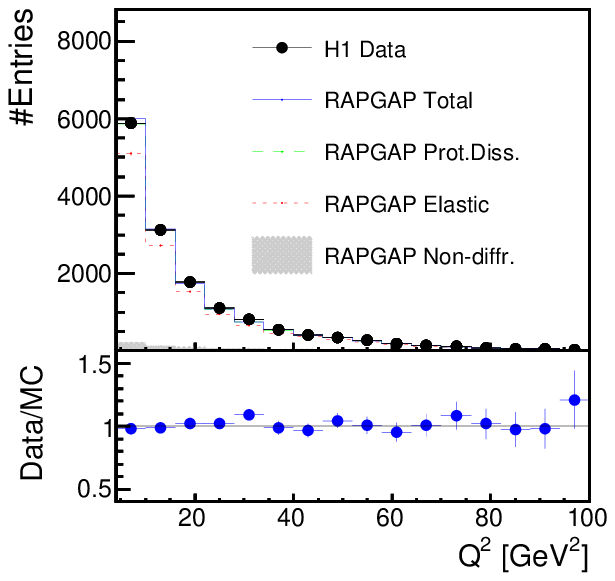}
}\hfill
\subfloat{
\includegraphics[width=.48\textwidth]{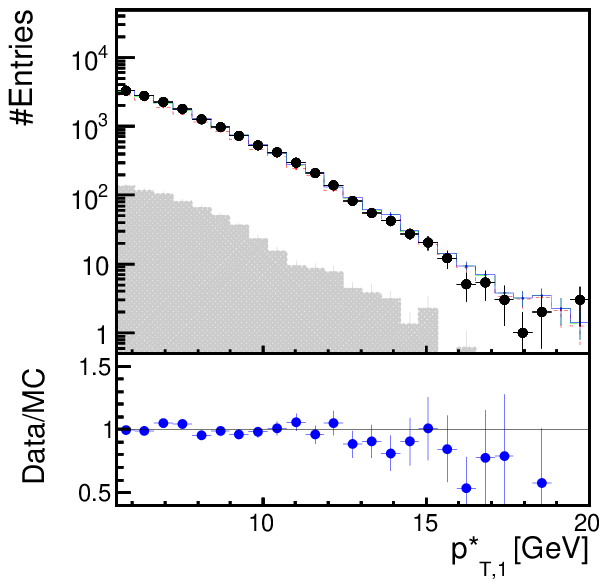}
}\hfill
\subfloat{
\includegraphics[width=.48\textwidth]{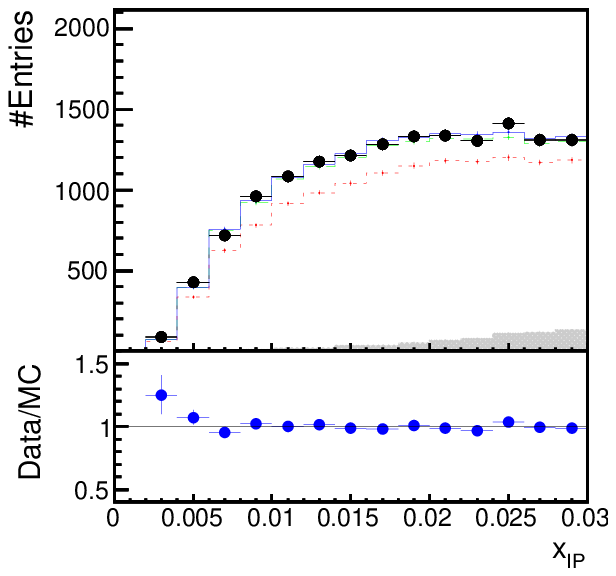}
}\hfill
\subfloat{
\includegraphics[width=.48\textwidth]{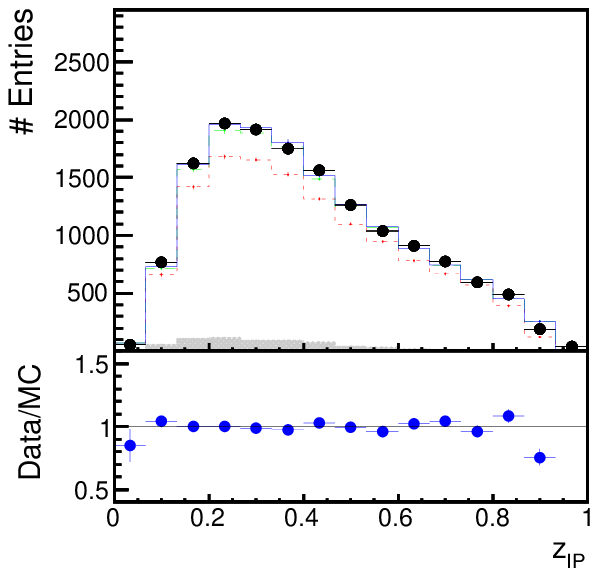}
}\hfill
\caption{Distributions of the kinematic quantities $\qsq$, $p^{\ast}_{\rm T,1}$, $\xpom$
		and $\zpom$. The data are shown as black points compared to the sum of MC simulation estimates.
		The filled area shows the contribution of non-diffractive DIS,
		the dotted line shows the diffractive contribution with the elastically scattered proton added to the non-diffractive DIS
		and the dashed line displays the proton dissociation contribution added to the diffractive contribution with the elastically 
		scattered proton and the non-diffractive DIS contribution.
		The sum of all contributions including the resolved photon processes is given by the full line.
		The MC is reweighted to the data. The ratio of data to the MC prediction is shown in the lower part of 
		of the individual figures.
		}
\label{fig:figctr}
\end{figure}

Cross sections at the level of stable hadrons are
obtained from the measured event rates in data
by applying corrections determined using the MC simulation.
In figure~\ref{fig:figctr} kinematic distributions
of the observables $\qsq$, $p^{\ast}_{\rm T,1}$, $\xpom$ and $\zpom$ as observed in the detector are shown
in comparison to the expectations from the reweighted MC simulation. The overall good description
of the data is achieved after applying 
a dedicated weighting of the MC simulation
in the variables $\zpom$, $\xpom$ and
 $x_{dijet}=
\sum_{1,2}(E^{\ast jet}-p^{\ast
  jet}_{z})_{i}/\sum_{HFS}(E^{\ast}-p^{\ast}_{z})_{i}$.
Weights are obtained from the reconstructed kinematic distributions 
and are applied at the hadron level.
This procedure is iterated until a good description of the shapes of the observables is achieved.

The data are corrected for detector inefficiencies, acceptance and 
finite resolution using the regularised unfolding procedure as implemented in TUnfold~\cite{unf_3}. 
A detector response matrix $A$, with elements $a_{ij}$ expressing the probability for an observable
originating in the generated MC sample from an interval $i$ to be measured in an interval $j$, 
is determined using the MC simulation. Migrations from outside the measurement 
phase space are included by additional rows of the detector response matrix.
The domains of jets with $3.0<p^{*}_{\rm T,1}<5.5\ \rm GeV$ and of events
with $0.03<\xpom<0.04$ are found to be the dominating sources of these migrations.
The MC simulation is reweighted in order to describe
the data also in these regions beyond the nominal phase space.

Two sources of background are considered in this analysis
and are subtracted from the data using Monte Carlo simulations prior to unfolding:
diffractive dijet events with $M_{Y}>1.6\ \rm GeV$ and $|t|<1\ \rm GeV^2$
and background from non-diffractive DIS.

For a background subtracted measurement $y_{j}$, the corresponding
number of events in the truth bin $i$, $x_{i}$,
is found by solving a minimisation problem 
for a $\chi^{2}$ function
\begin{equation}
\chi^{2}=(y-Ax)^{T}V_{yy}^{-1}(y-Ax)+\tau^{2}x^{2}, 
\label{eq:unf}
\end{equation} 
where $x$ and $y$ are vectors defined by $y_{j}$ and $x_{i}$, respectively,
$V_{yy}$ is the covariance matrix accounting for the statistical uncertainties of $y_{j}$ and 
$\tau$ is a regularisation parameter introduced in order to damp statistical fluctuations of the solution.
The regularisation parameter $\tau$ is determined using the L-Curve scan~\cite{unf_3}.

The cross section 
in each measurement interval $i$ is given by
\begin{equation}
\sigma_{i}(ep\to ep'X)=\frac{x_{i}}{\cal L}(1+\delta_{i, rad}),
\end{equation}
where $\cal L$ is the integrated luminosity of the data sample and $(1+\delta_{i, rad})$
is the correction for QED radiation effects in the interval $i$.   
These corrections are calculated as a ratio of RAPGAP predictions
with and without QED radiation simulated.
The differential cross section is determined by dividing $\sigma_{i}$ by the area of the corresponding interval.

\subsection{Systematic uncertainties}

The systematic uncertainties induced by experimental effects
and by model adequateness are propagated to each measurement interval 
in the unfolding procedure (eq.~\ref{eq:unf}). A dedicated detector response matrix is constructed
for each variation related to particular sources of uncertainties:  
\begin{itemize}
\item{The energy of the scattered electron is varied by $\pm 1\%$ with a resulting uncertainty on the integrated 
dijet cross section of $1\%.$}
\item{The polar angle of the scattered electron is varied by $\pm 1\ \rm mrad$ with a resulting uncertainty 
on the integrated dijet cross section of $1\%.$}
\item{The energy of each particle contained in HFS is varied by $\pm 1\%$~\cite{kogler} which translates into an uncertainty 
on the integrated dijet cross section of $4\%$.}
\item{
Uncertainties related to the model dependent corrections of the data
are accounted for by varying the shape of the kinematic distributions in
$\qsq$, $\xpom$, $\beta$, $p^{\ast}_{\rm T,1}$, $\zpom$, $x_{dijet}$ and $\Delta\eta^{\ast}$
in the MC such that the data are still described  within the statistical uncertainties.
For this purpose, the multiplicative weights
$(\log\qsq)^{\pm 0.2}$, $\xpom^{\pm 0.05}$, $\beta^{\pm 0.01}(1-\beta^{\pm 0.01})$, $p^{\ast \pm 0.04}_{\rm T,1}$, $\zpom^{\pm 0.15}$,
$x^{\pm 0.15}_{dijet}$ and $(1.5+\Delta\eta^{\ast})^{\pm 0.5}$ are applied, respectively.
The largest resulting uncertainty of $3\%$ arises from the variation of the shape in $p^{\ast}_{\rm T,1}$.
The shape of the distribution in $t$ is varied within the experimental uncertainty on the $t$-slope~\cite{tslope}
by applying a weight of $e^{\pm t}$ in MC, which translates into an uncertainty on the integrated dijet cross section of $1\,\%$.
The integrated cross section uncertainty due to the model dependence of the measurement is of the order of $5\,\%$.
}
\end{itemize}
 
The following uncertainties on the global normalisation are considered:
\begin{itemize}
\item{The luminosity of the data is measured with a precision $\pm 2.7\ \%$~\cite{qedclumi}.}
\item{The trigger efficiency related to the tracking and  SpaCal condition 
	induces an uncertainty of $2\%$ and $1\%$, respectively.}
\item{The uncertainty accounting for the LRG selection efficiency is $7\%$~\cite{incldiff_lrg}.}
\item The normalisation of the non-diffractive DIS background modelled by RAPGAP is varied by $\pm 50\,\%$
and the normalisation of the diffractive background is varied by $\pm 100\,\%$,
yielding a resulting uncertainty on the integrated dijet cross section below $1\%$ in both cases.

\end{itemize}

The total systematic uncertainty is obtained by adding the individual contributions in quadrature.

\section{Results}


The integrated cross section in the measurement phase space specified in table~\ref{tab:ps} is found to be
\begin{equation}
\sigma^{dijet}_{meas}(ep\to eXY) = 73\pm{2}\ (\rm stat.)\ \pm{7}\ (\rm syst.)\ \rm pb~.
\end{equation}
The NLO QCD prediction of the total diffractive dijet cross section is
\begin{equation}
\sigma^{dijet}_{theo}(ep\to eXY) = 77\ ^{+25}_{-20}\ (\rm scale)\ ^{+4}_{-14}\ (\rm DPDF)\ \pm{3}\ (\rm had)\ \rm pb~,
\end{equation}
in very good agreement with the measurement.
The uncertainty on the NLO prediction is found to be significantly larger than the experimental uncertainty.

\begin{figure}[b]
\subfloat{
\includegraphics[bb=80bp  100bp 380bp 400bp, width=.4\textwidth]{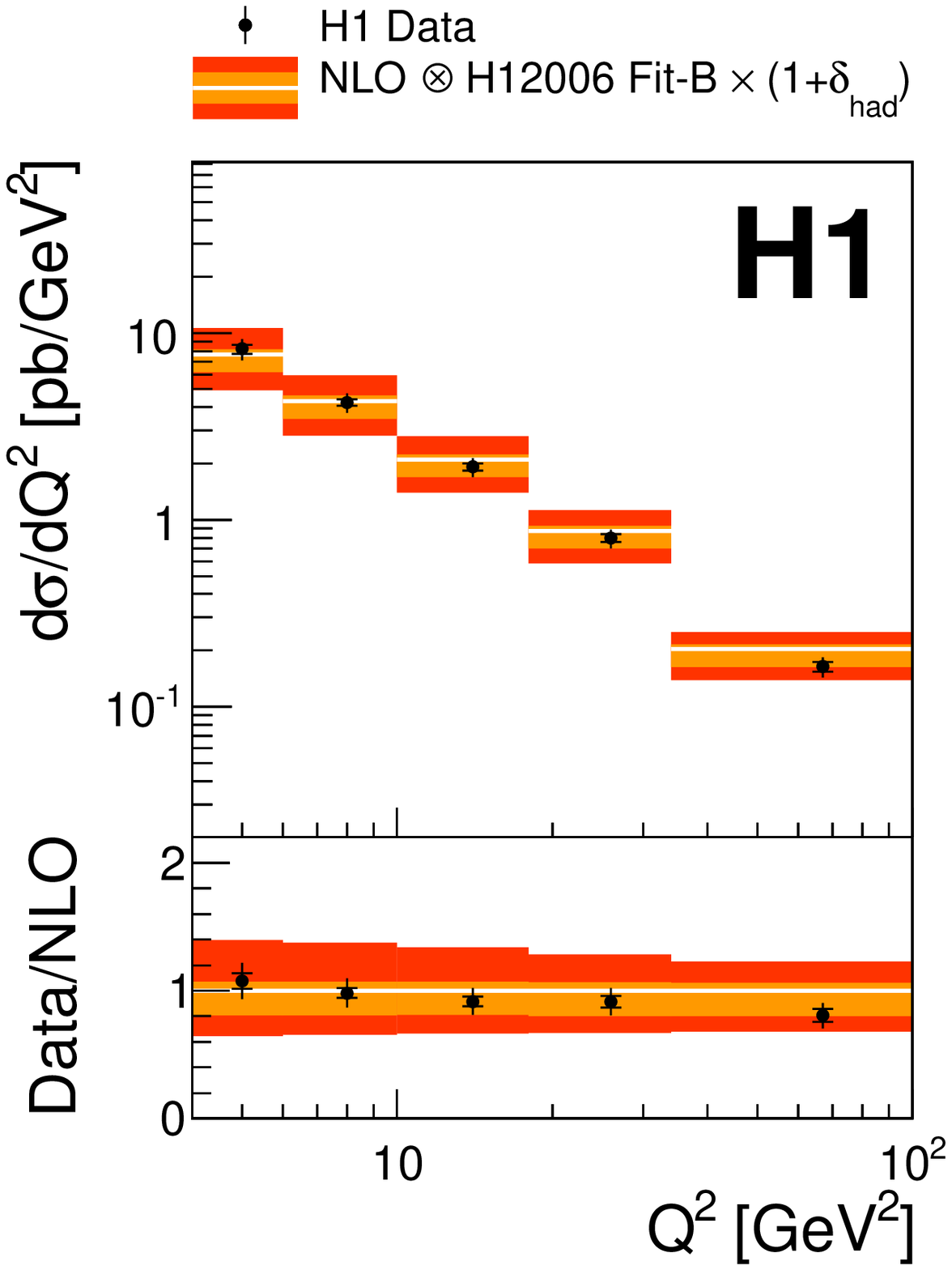}
}\hfill
\subfloat{
\includegraphics[bb=30bp  100bp 330bp 400bp, width=.4\textwidth]{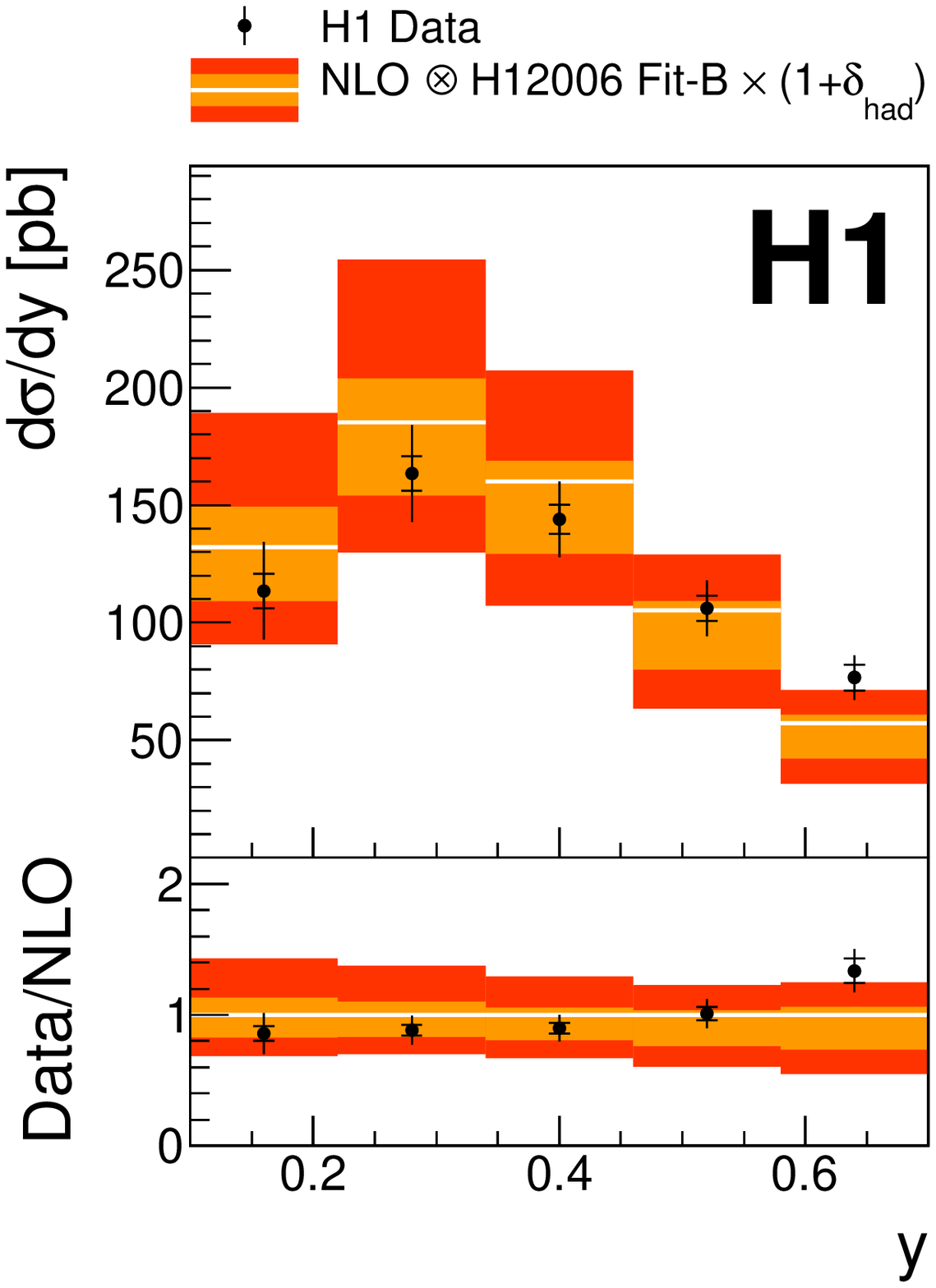}
}\hfill
\vspace{2cm}
\caption{Diffractive dijet differential cross section as a function of $\qsq$ and $y$.
        The inner error bars on the data points represent the statistical uncertainties, while the outer error bars include
        the systematic uncertainties added in quadrature. The NLO QCD prediction based on the H12006 Fit-B DPDF set is displayed as a white line. 
        The light shaded band indicates the uncertainty arising from hadronisation and the DPDF fit added
        in quadrature. The outer dark band shows the full theory uncertainty including the QCD scale uncertainty added in quadrature. 
	The ratio of the single-differential cross section to the NLO prediction is shown in the lower part of 
	the individual figures.
	}
\label{fig:fig1}
\end{figure}

\begin{figure}[b]
\subfloat{
\includegraphics[bb=80bp  100bp 380bp 400bp, width=.4\textwidth]{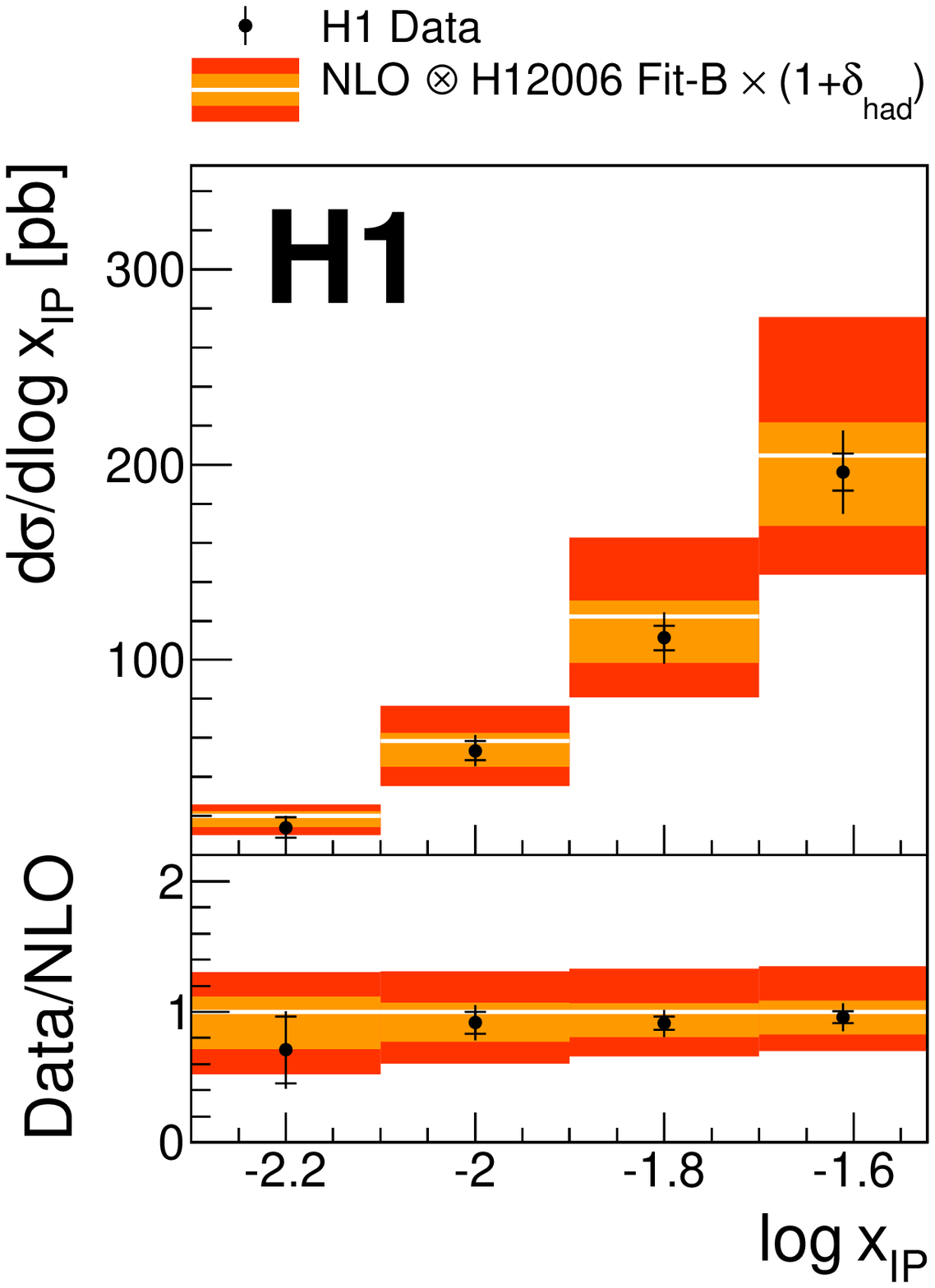}
}\hfill
\subfloat{
\includegraphics[bb=30bp  100bp 330bp 400bp, width=.4\textwidth]{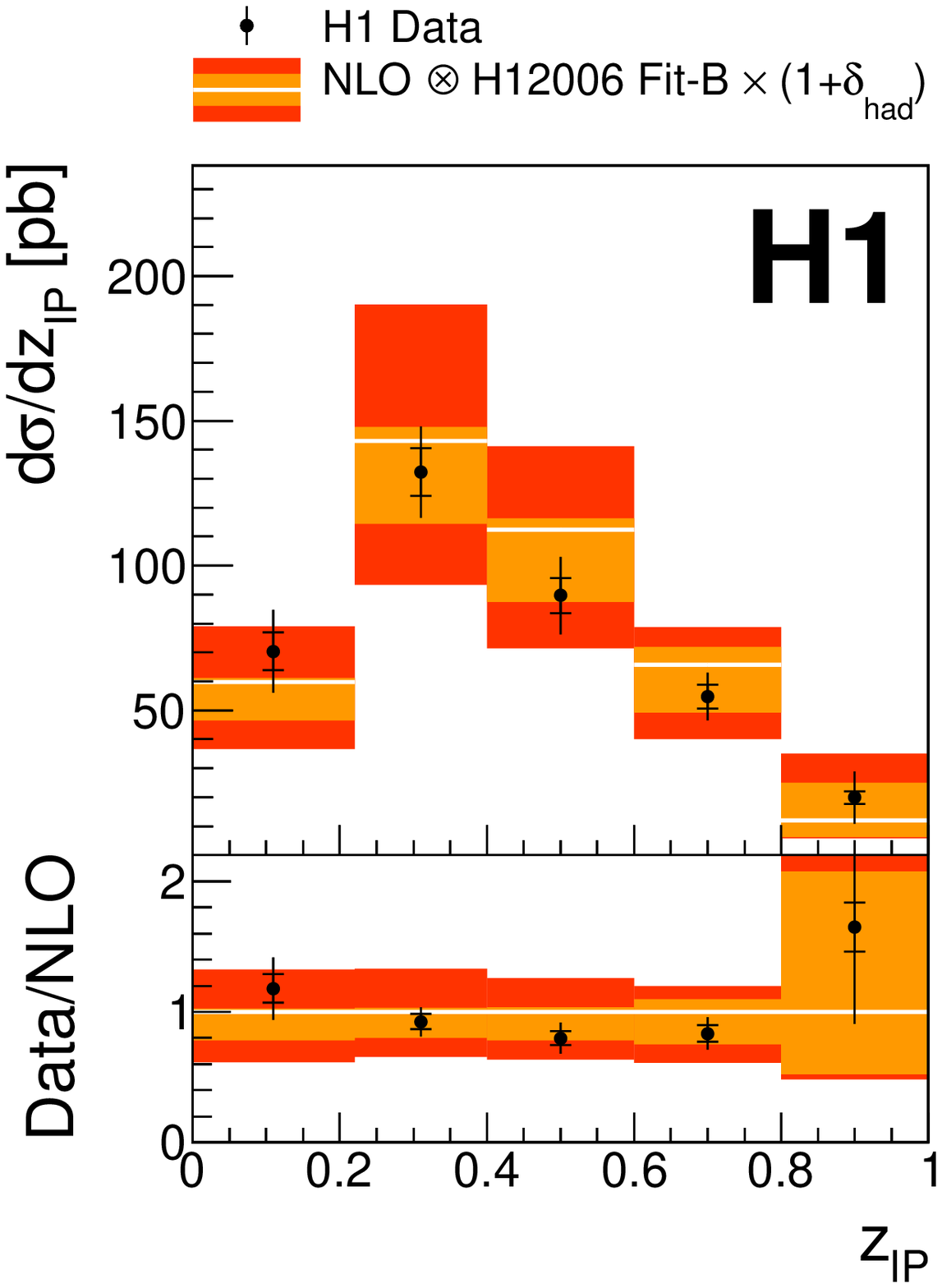}
}\hfill
\vspace{2cm}
\caption{Diffractive dijet differential cross section as a function of $\log\xpom$ and $\zpom$.
        The inner error bars on the data points represent the statistical uncertainties, while the outer error bars include
        the systematic uncertainties added in quadrature.
	Further details are given in figure~\ref{fig:fig1}.} 
\label{fig:fig2}
\end{figure}

\begin{figure}[b]
\subfloat{
\includegraphics[bb=80bp  100bp 380bp 400bp, width=.4\textwidth]{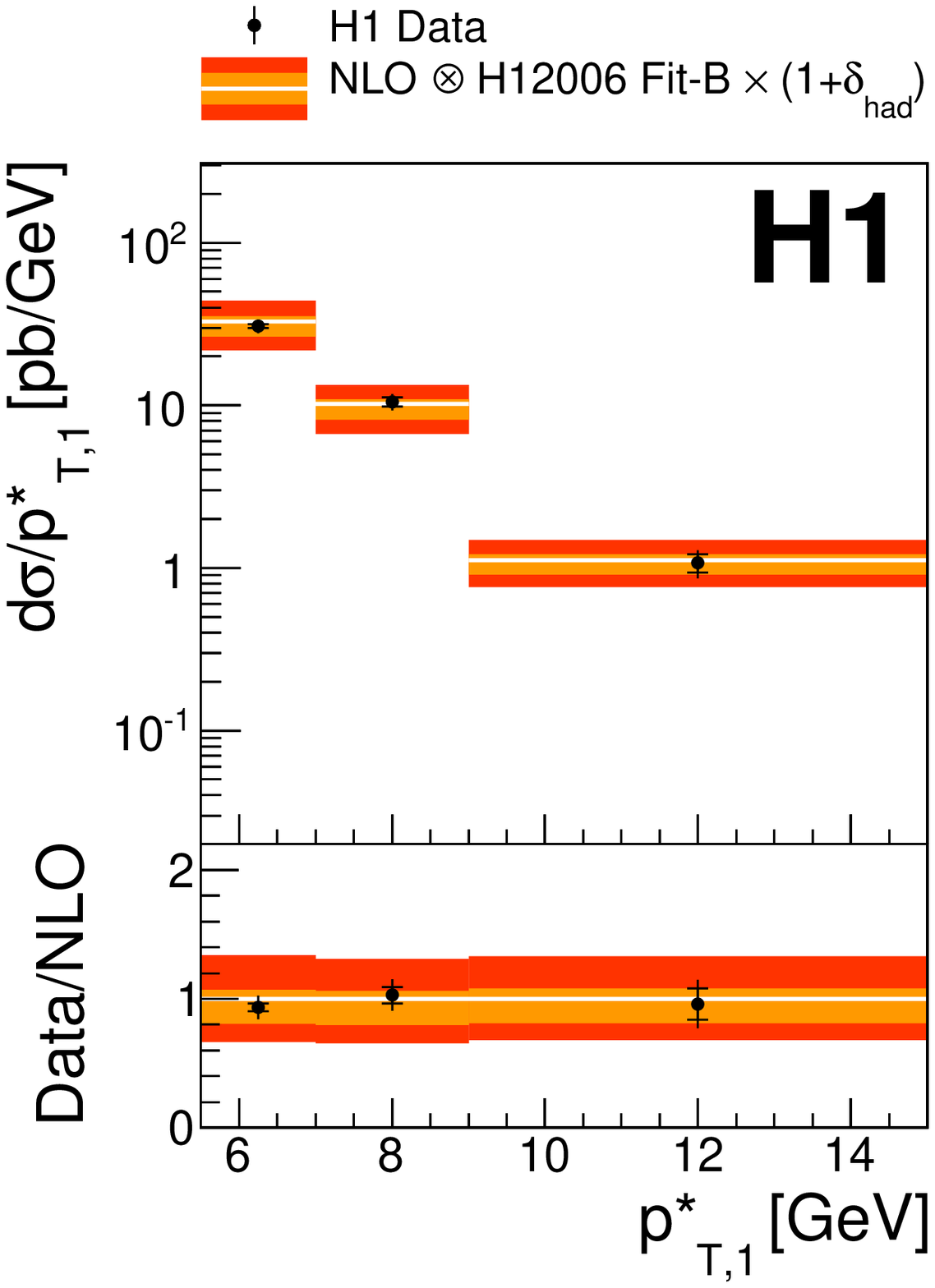}
}\hfill
\subfloat{
\includegraphics[bb=30bp  100bp 330bp 400bp, width=.4\textwidth]{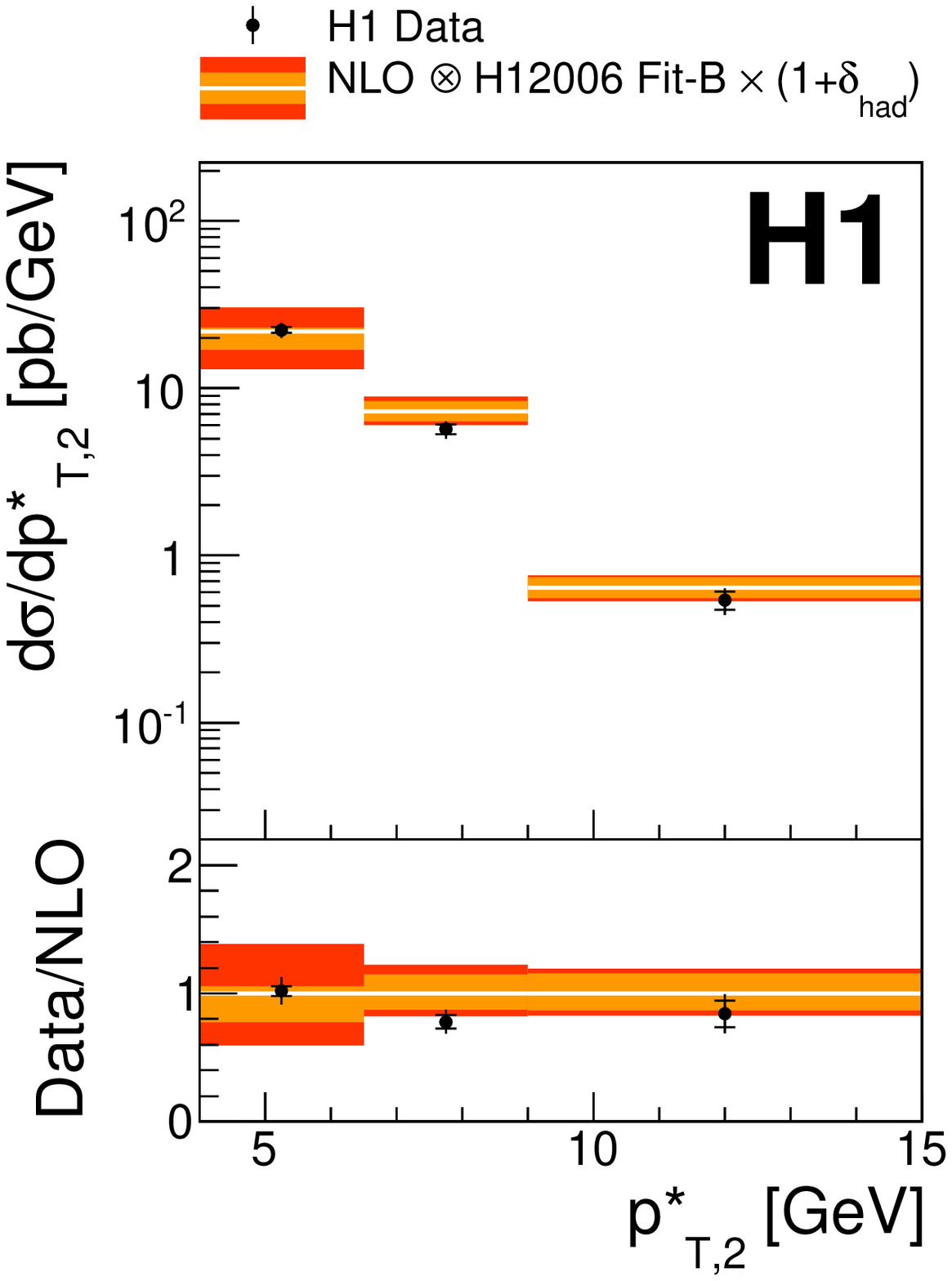}
}\hfill
\vspace{2cm}
\caption{Diffractive dijet differential cross section as a function of $p^{\ast}_{\rm T,1}$ and $p^{\ast}_{\rm T,2}$.
        The inner error bars on the data points represent the statistical uncertainties, while the outer error bars include
        the systematic uncertainties added in quadrature. 
        Further details are given in figure~\ref{fig:fig1}.}
\label{fig:fig3}
\end{figure}

\begin{figure}[b]
\subfloat{
\includegraphics[bb=80bp  100bp 380bp 400bp, width=.4\textwidth]{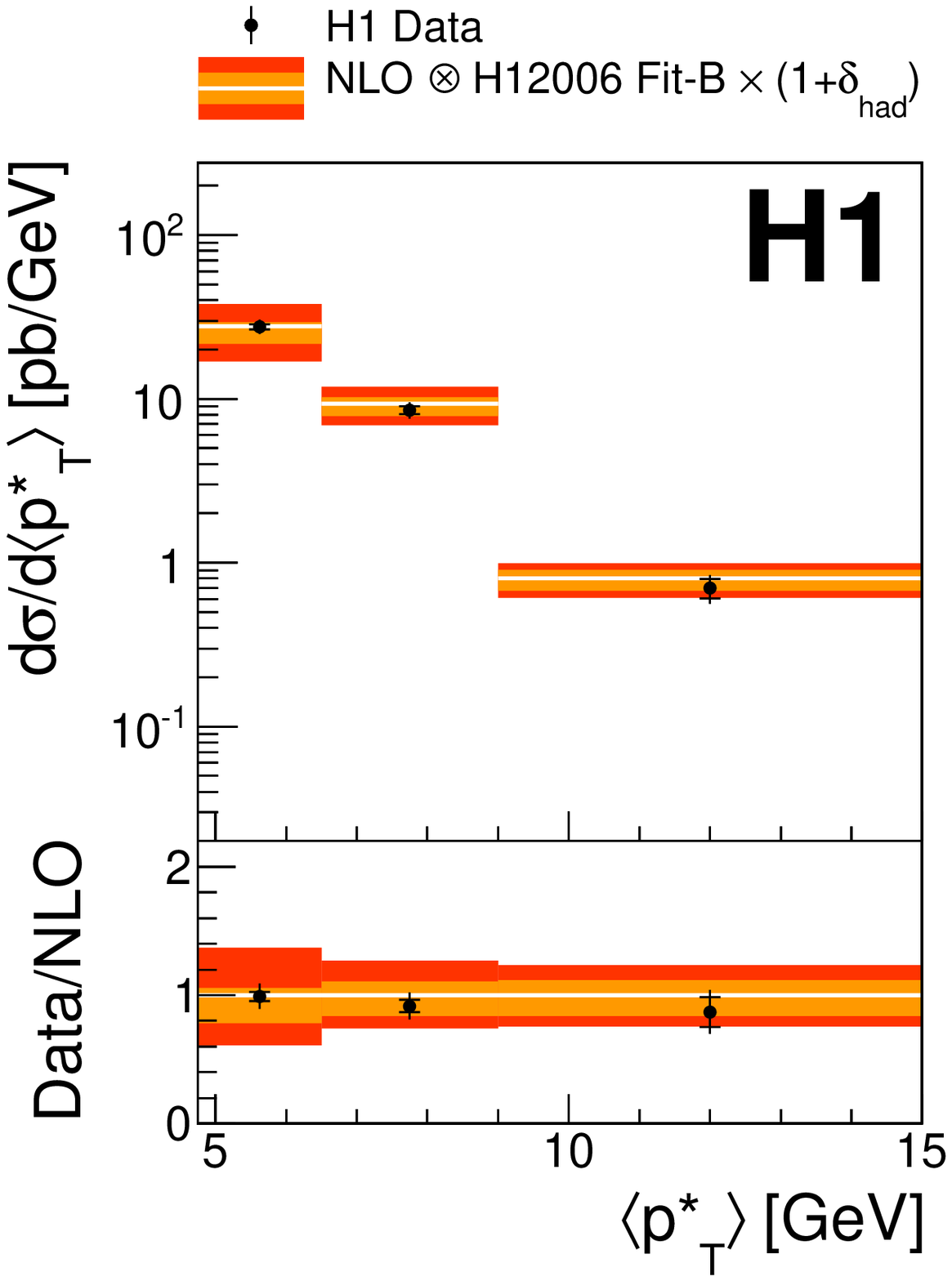}
}\hfill
\subfloat{
\includegraphics[bb=30bp  100bp 330bp 400bp, width=.4\textwidth]{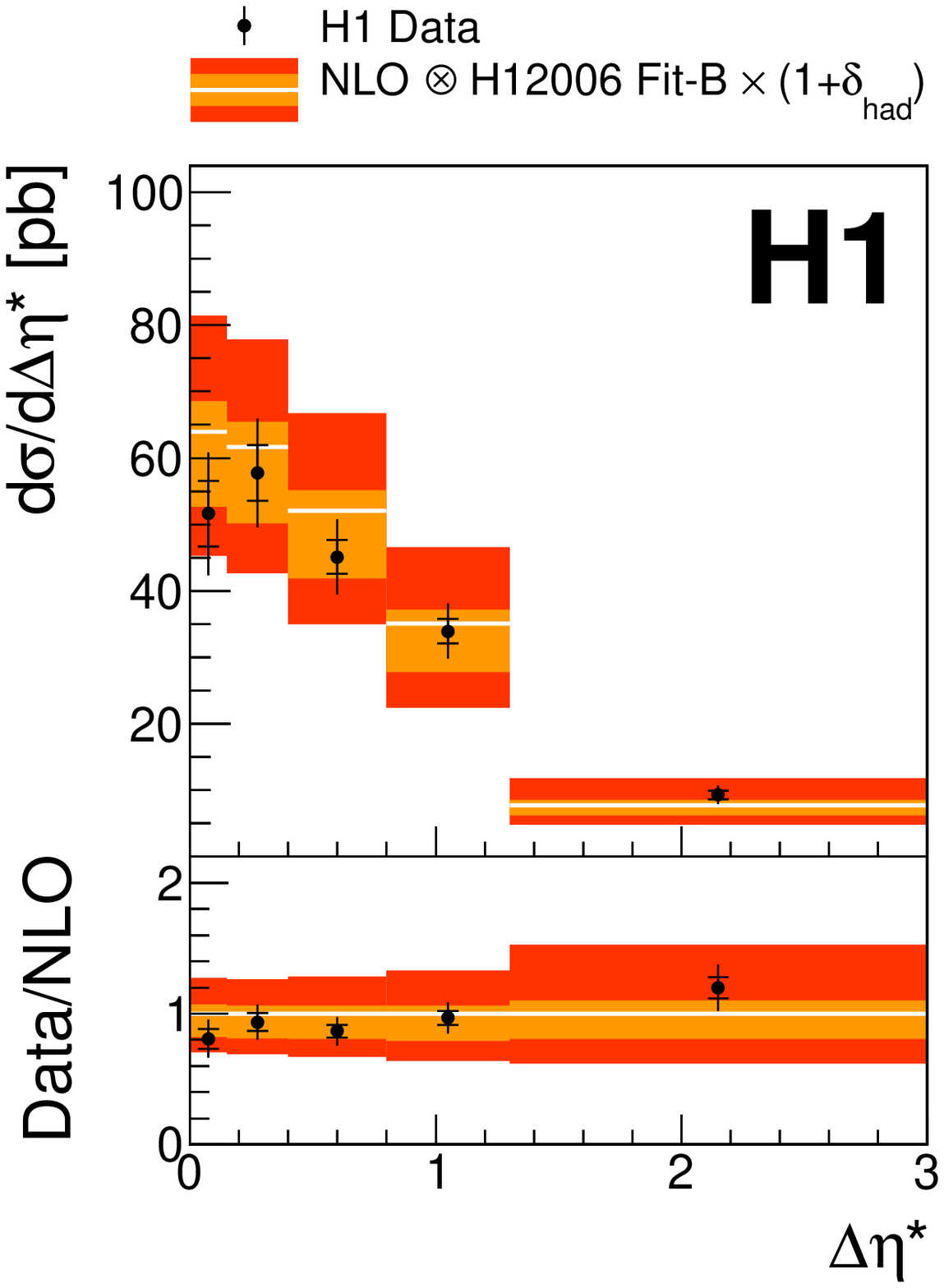}
}\hfill
\vspace{2cm}
\caption{Diffractive dijet differential cross section as a function of $\langle p^{\ast}_{\rm T}\rangle$ and $\Delta\eta^{\ast}$.
        The inner error bars on the data points represent the statistical uncertainties, while the outer error bars include
        the systematic uncertainties added in quadrature. 
        Further details are given in figure~\ref{fig:fig1}.}
\label{fig:fig4}
\end{figure}

Single differential cross sections are given in tables~\ref{tab:tab1} and~\ref{tab:tab2} and are shown in 
figures~\ref{fig:fig1}-\ref{fig:fig4}. The statistical correlations between measurements
in different bins are given in tables~\ref{tab:tabrho1} and~\ref{tab:tabrho2}.  
The differential cross sections as a function of the DIS variables $\qsq$ and $y$
are shown in figure~\ref{fig:fig1}, as a function of the momentum fractions $\xpom$ and $\zpom$
are shown in figure~\ref{fig:fig2} and as a function of the jet variables $p^{\ast}_{T,1}$, 
$p^{\ast}_{T,2}$, $\langle p^{\ast}_{T}\rangle$ and $\Delta\eta^{\ast}$ 
are shown in Figure~\ref{fig:fig3} and~\ref{fig:fig4}.
For the majority of the measurements, the data precision is limited by systematic effects.
The statistical correlations are small for the inclusive kinematic variables $\qsq$ and $y$
and moderate ($|\rho|< 0.6$) for the other variables.
The figures also include the NLO QCD predictions
which describe within their large uncertainties the data well.

\begin{figure}[b!]
\includegraphics[bb=100bp  140bp 440bp 440bp, width=.64\textwidth]{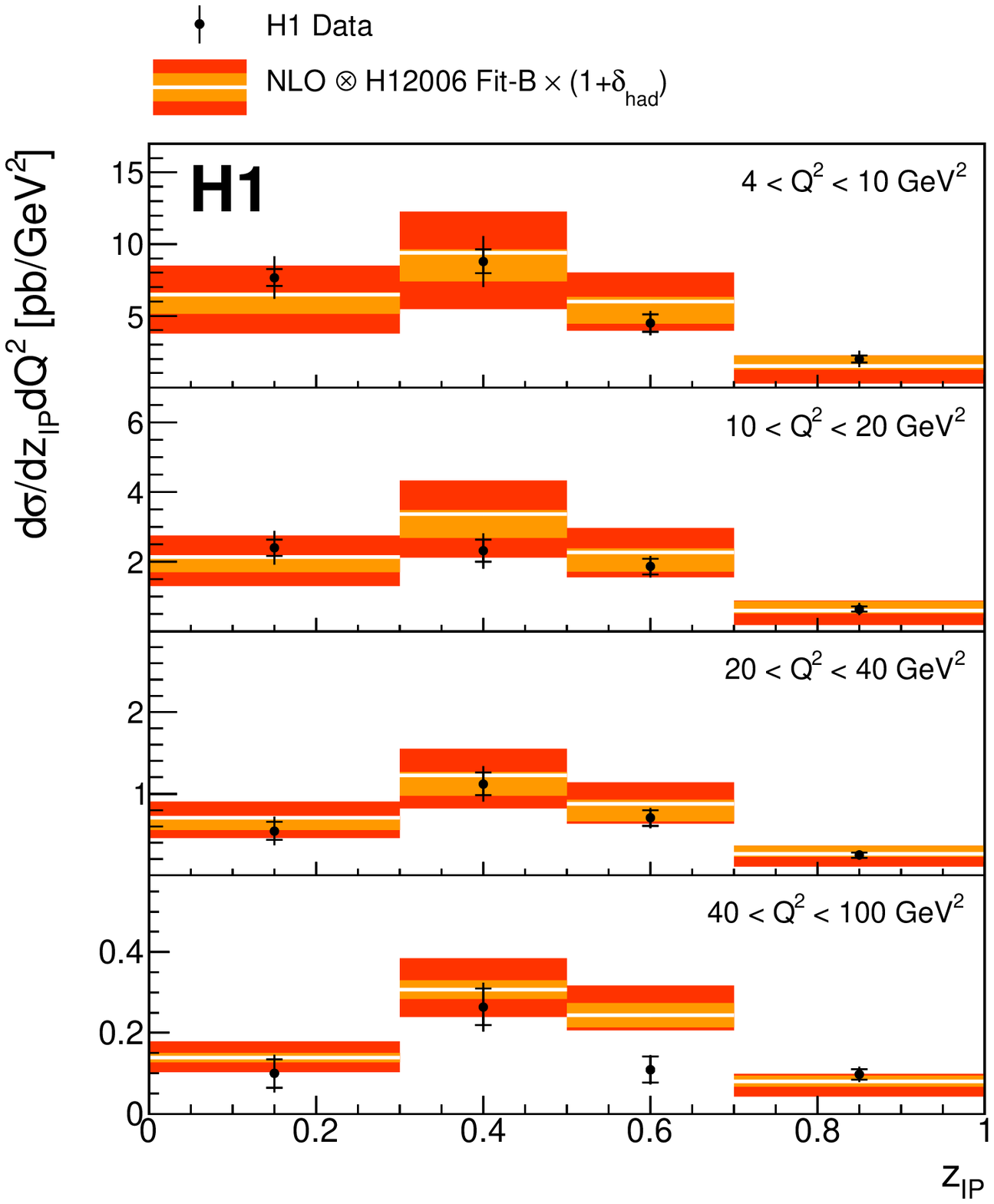}
\vspace{4cm}
\caption{Double-differential cross section as a function of $\zpom$ and $\qsq$.
        The inner error bars on the data points represent the statistical uncertainties, while the outer error bars include
        the systematic uncertainties added in quadrature. 
        Further details are given in figure~\ref{fig:fig1}.}
\label{fig:fig5}
\end{figure}

\begin{figure}[!b]
\includegraphics[bb=100bp  140bp 440bp 440bp, width=.64\textwidth]{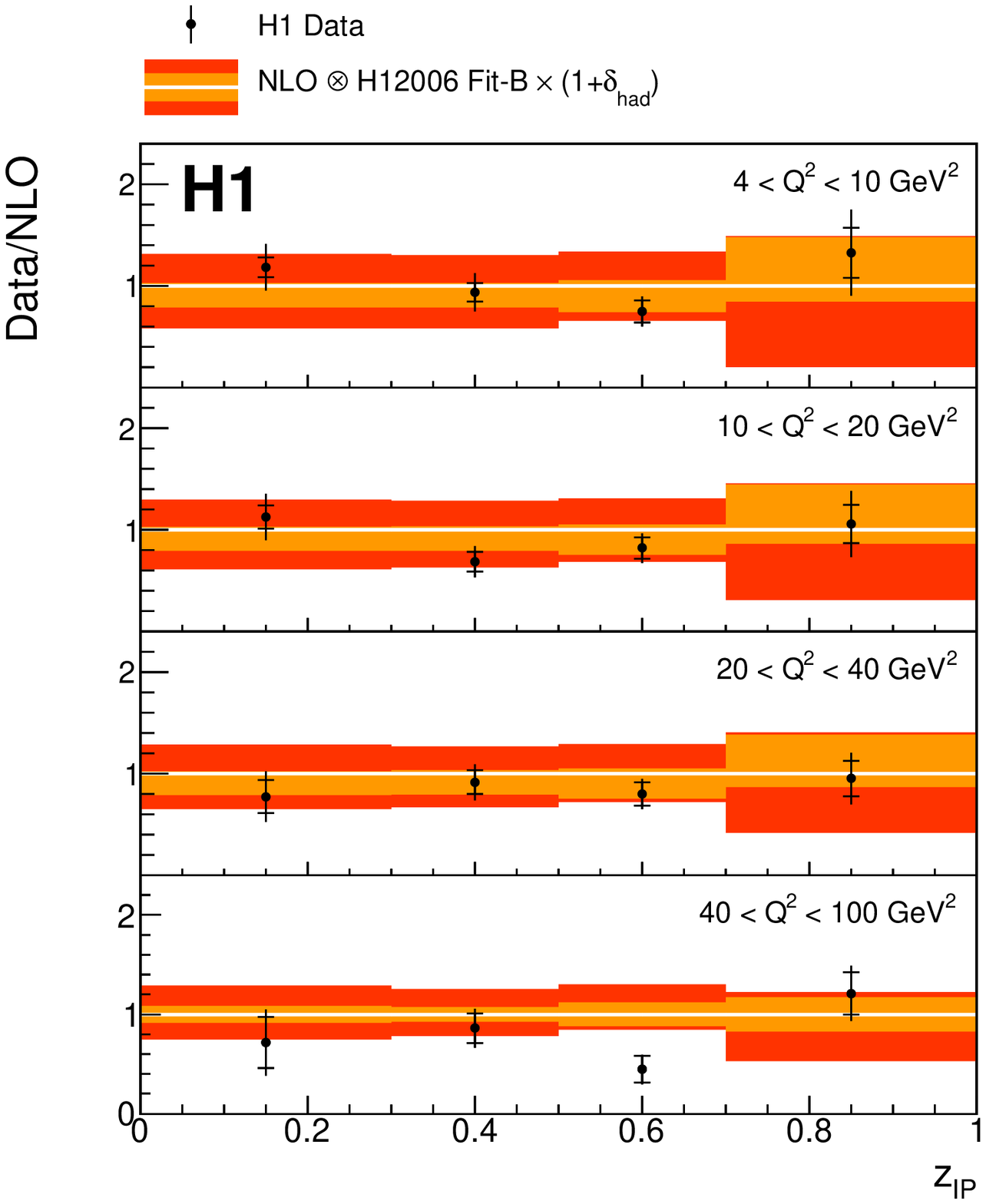}
\vspace{4cm}
\caption{Ratio of the double-differential cross section to the NLO prediction as a function of $\zpom$  and $\qsq$.
        The inner error bars on the data points represent the statistical uncertainties, while the outer error bars include
        the systematic uncertainties added in quadrature. 
        Further details are given in figure~\ref{fig:fig1}.}
\label{fig:fig6}
\end{figure}

\begin{figure}[!b]
\includegraphics[bb=40bp  180bp 380bp 480bp, width=.64\textwidth]{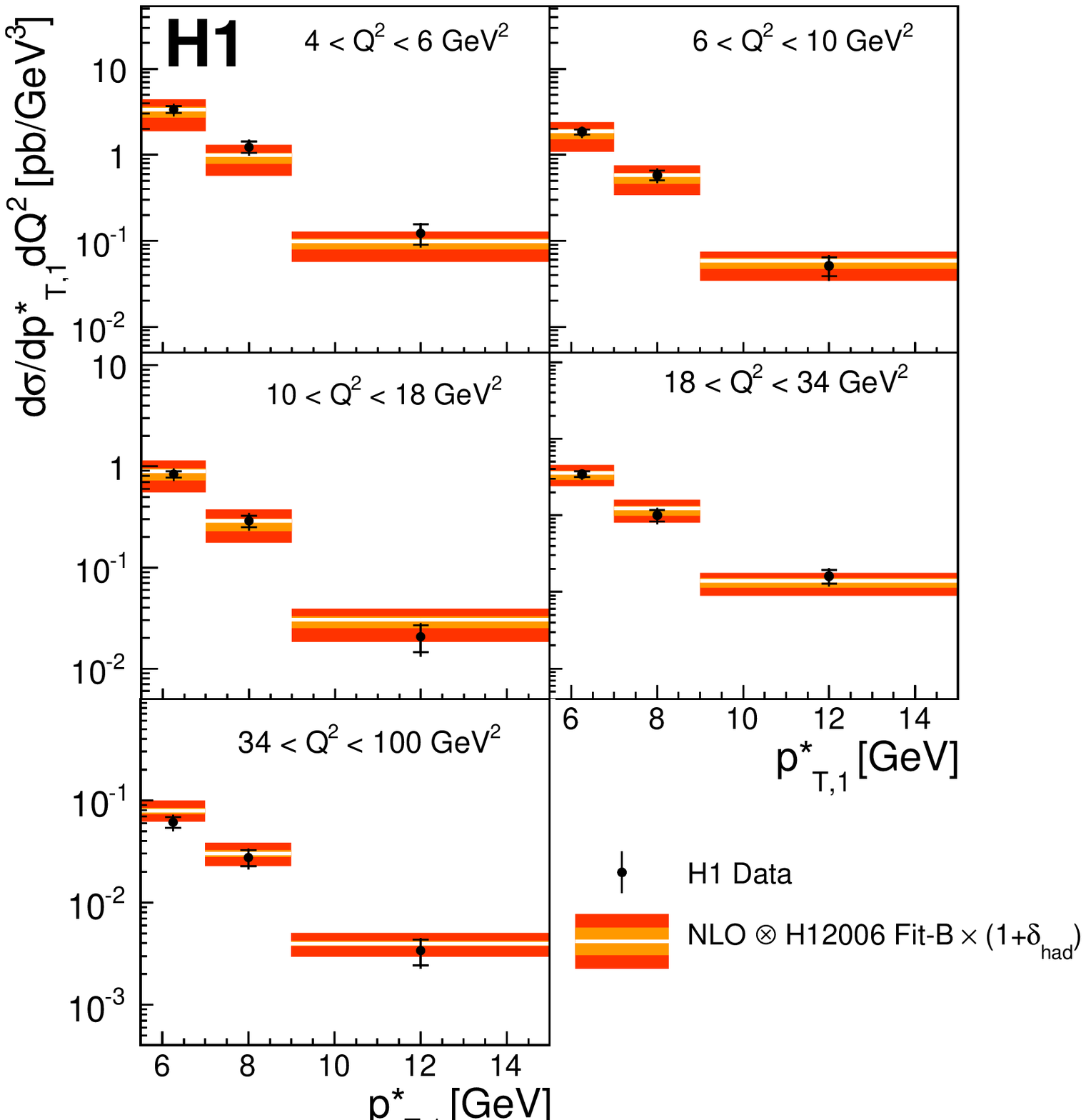}
\vspace{6cm}
\caption{Double-differential cross section as a function of $p^{\ast}_{\rm T,1}$ and $\qsq$.
        The inner error bars on the data points represent the statistical uncertainties, while the outer error bars include
        the systematic uncertainties added in quadrature. 
        Further details are given in figure~\ref{fig:fig1}.}
\label{fig:fig9}
\end{figure}

\begin{figure}[!b]
\includegraphics[bb=40bp  180bp 380bp 480bp, width=.64\textwidth]{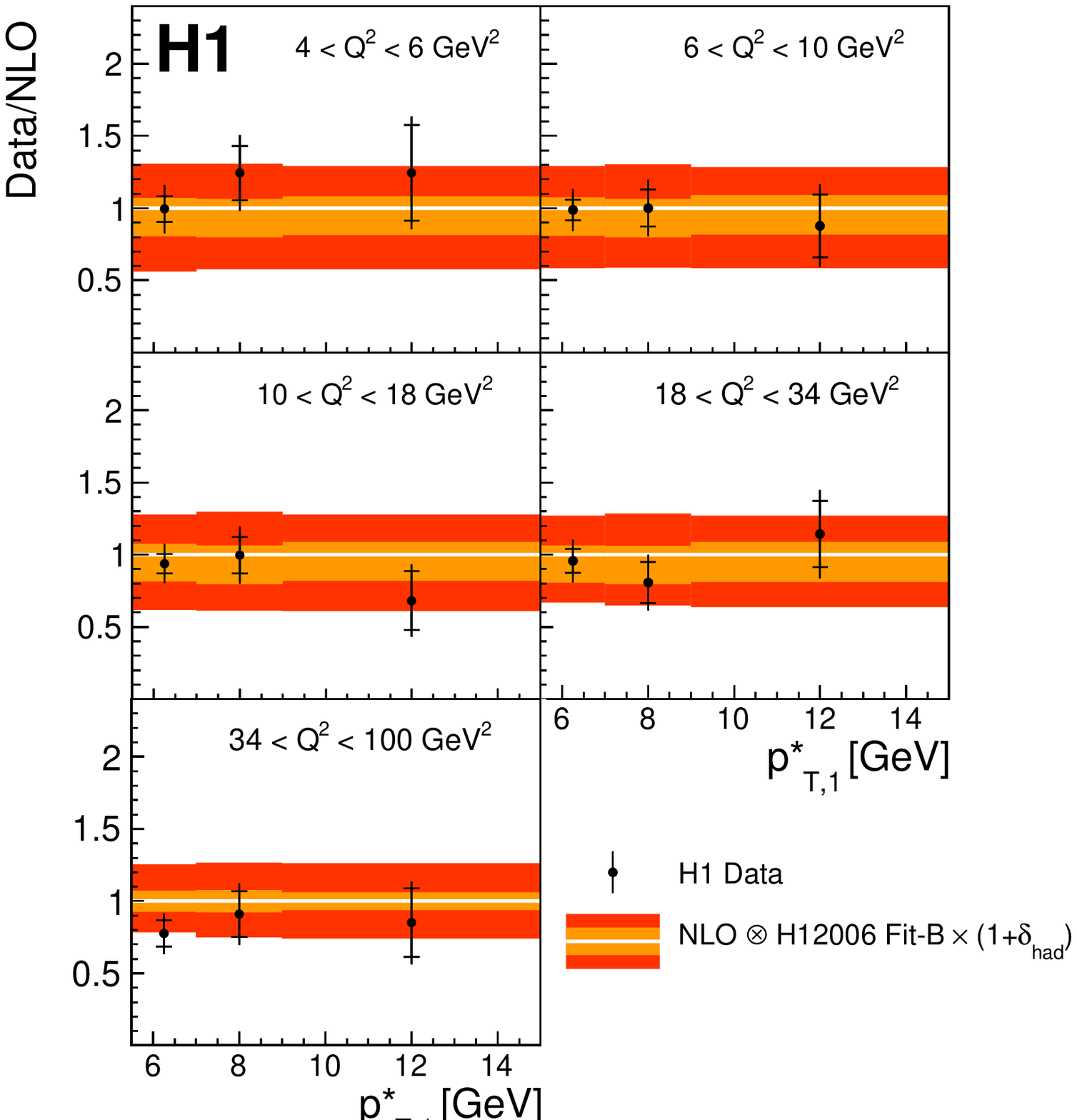}
\vspace{6cm}
\caption{Ratio of the double-differential cross section to the NLO prediction as a function of $p^{\ast}_{\rm T,1}$ and $\qsq$.
        The inner error bars on the data points represent the statistical uncertainties, while the outer error bars include
        the systematic uncertainties added in quadrature. 
        Further details are given in figure~\ref{fig:fig1}.}
\label{fig:fig10}
\end{figure}

The dynamics of dijet production is further studied in terms of double differential cross sections
in bins of $\zpom$ and of the QCD scale defining observables $\qsq$ and $p^{\ast}_{\rm T,1}$.
The double differential cross sections are listed in tables~\ref{tab:tab3}-\ref{tab:tab5}
and are shown in figures~\ref{fig:fig5}-\ref{fig:fig10}.
The corresponding statistical correlations between measurements
in different bins are given in tables~\ref{tab:tabrho3}-\ref{tab:tabrho5}.
Figure~\ref{fig:fig5} shows the double differential cross section measured in bins of $\zpom$ and $\qsq$.
The ratio of the data to the theory prediction is shown in figure~\ref{fig:fig6}. 
The data are well described by the NLO prediction in most of the phase space. 
The double differential cross section measured in bins of $p^{\ast}_{\rm T,1}$ and $\qsq$ is shown in figure~\ref{fig:fig9}
and the corresponding ratios of the measurements to the NLO predictions are shown in figure~\ref{fig:fig10}.

The present measurement is based on a 
six times increased
luminosity as compared to the previous H1 measurement of dijet
production with LRG~\cite{jetdiff_3} and is using a more sophisticated data
correction method. 
A direct comparison of the present data to other measurements of dijet production in diffractive DIS is not possible because of different phase space definitions.
Measurements based on the direct detection of a
forward proton~\cite{jetdiff_2} are limited in statistical precision due
to the restricted geometrical acceptance of the proton taggers.

The experimental uncertainties on both single- and double-differential cross sections
are in general smaller than the theory uncertainties.
The data thus have the power to constrain QCD in diffractive DIS.
Here, the double-differential dijet cross sections as a function of $\qsq$ and $p_{\rm T,1}^*$ are used 
to determine the value of the strong coupling constant $\asmz$ at the scale 
of the mass of the $Z$-boson, $M_Z$. 
The value of $\asmz$ is determined by an iterative $\chi^2$-minimisation 
procedure using NLO calculations, corrected for hadronisation effects
following the method~\cite{HighQ2Paper}.
In the fit, the uncertainties on the HFS energy scale are treated as $50\%$ correlated and $50\%$ uncorrelated. 
All other experimental uncertainties are treated as correlated.
Scale uncertainties, hadronisation uncertainties and  DPDF uncertainties of the NLO calculation
are propagated to the fit result as described in~\cite{HighQ2Paper}.

The fit yields a value of $\chi^2/n_{\rm dof} = 16.7/14$, with $n_{\rm dof}$ being the number of degrees of freedom,
thus indicating good agreement 
of theory to data. The nuisance parameters of the correlated systematic uncertainties 
are equally distributed around zero with absolute values below one.
The value of $\asmz$ determined in the fit to the dijet cross sections is
\begin{flalign}
 \asmz &= 0.119\pm0.004\,({\rm exp})\pm0.002\,({\rm had})\pm0.005\,({\rm DPDF})\pm0.010\,({\mu_r})\pm0.004\,({\mu_f}) \\
       & =0.119\pm0.004\,({\rm exp})\pm0.012\,({\rm DPDF,theo}) \nonumber
\end{flalign}
The largest uncertainties arise from the estimate of the contributions from  orders beyond NLO
and from the poor knowledge of the DPDF.
The largest contribution to the experimental uncertainty of $0.003$
arises from the global normalisation uncertainty.

The result for $\asmz$ is consistent within the uncertainties with the world average
~\cite{Beringer:1900zz,Bethke:2012jm} and with values from other jet data 
in DIS and photoproduction~\cite{Aaron:10:1,HighQ2Paper,Abramowicz:2012jz} as well as values 
of $\asmz$ determined from jet data at 
the Tevatron~\cite{Abazov:2009nc,Abazov:2012lua} and at the LHC~\cite{Chatrchyan:2013txa,alps_cms}.
Although the uncertainty of this $\asmz$ extraction is not competitive
with measurements in other processes the agreement with the other
measurements supports the underlying concept of treating dijet
production in diffractive DIS with perturbative QCD calculations. 
\section{Conclusions}

Integrated, single- and double-differential cross sections of diffractive DIS dijet production
are measured with the H1 experiment in $ep$ collisions at HERA
and compared with NLO QCD predictions.

The integrated diffractive dijet cross section is found to be well
described by the NLO QCD prediction using the H12006 Fit-B DPDF set.
Both shapes and normalisation of the single-differential cross sections are 
reproduced by the theory within the experimental and theory uncertainties,
confirming at improved precision the conclusions from previous H1 measurements. 
Good agreement of the theory with the measurement is also
found for the shapes and normalisation of the double differential cross sections.
The cross section measurements presented here show 
experimental uncertainties significantly smaller than the uncertainties of the
theory predictions. 
From a fit of the NLO prediction to the double differential cross sections in $\qsq$ and $p^{*}_{T,1}$,  
the strong coupling constant has been determined to be $\alpha_{s}(M_{Z})=0.119\ (4)_{\rm exp}\ (12)_{\rm theo}$.

\section*{Acknowledgements}
We are grateful to the HERA machine group whose outstanding efforts have made this 
experiment possible. We thank the engineers and technicians 
for their work in constructing and
maintaining the H1 detector, our funding agencies for 
financial support, the DESY technical
staff for continual assistance and the DESY directorate for
the hospitality which they extend to
the non-DESY members of the collaboration.
We would like to give credit to all partners contributing to the EGI computing infrastructure for their support for the H1 Collaboration.

\begin{flushleft}

\end{flushleft}

\newpage
\clearpage
\begin{landscape}
\begin{table}
\resizebox{.8\textwidth}{!}{\begin{minipage}{\textwidth}
\begin{tabular}{|c|c|c|c|c||c|c|c|c|c|c|c|c|c|c|c|c||c||c|}
\hline
$\qsq$ & $d\sigma/d\qsq$ & $\delta_{tot}$ & $\delta_{stat}$ & $\delta_{sys}$ & $\delta_{\theta}$ & $\delta_{E}$ & $\delta_{HFS}$
        & $\delta_{\qsq}$ & $\delta_{\xpom}$ & $\delta_{\beta}$ & $\delta_{p*_{T,1}}$ 
        & $\delta_{\zpom}$ & $\delta_{x_{dijet}}$ & $\delta_{\Delta\eta*}$ & $\delta_{t}$
        & $\delta_{bgr}$ & $1+\delta_{had}$ & $1+\delta_{rad}$\\
$\rm [GeV^2]$ & $\rm [pb/GeV^2]$ & $[\%]$ & $[\%]$ & $[\%]$ & $[\%]$ & $[\%]$ & $[\%]$ & $[\%]$ &  $[\%]$ &  $[\%]$ &  $[\%]$ &  $[\%]$ &
        $[\%]$ &  $[\%]$ &  $[\%]$ & $[\%]$ & & \\
\hline
$4 \div 6$ &$8.20$ &$13.2$ & $5.7$ & $11.9$ &$1.0$ & $4.5$ & $-3.9$ &$2.1$ & $1.1$ &$2.8$ & $-5.0$ &$1.9$ & $0.7$ &$1.2$ & $-0.9$ &$0.1$ &$1.05 \pm 0.05$ &$1.05$ \\
$6 \div 10$ &$4.23$ &$11.8$ & $4.0$ & $11.0$ &$2.6$ & $1.7$ & $-5.0$ &$-0.5$ & $0.5$ &$-0.5$ & $-3.1$ &$-3.2$ & $-1.6$ &$1.6$ & $-1.5$ &$0.3$ &$1.05 \pm 0.04$ &$1.03$ \\
$10 \div 18$ &$1.92$ &$11.4$ & $4.0$ & $10.7$ &$1.0$ & $1.9$ & $-4.6$ &$-0.9$ & $0.6$ &$-0.8$ & $-3.1$ &$-3.1$ & $-1.5$ &$1.7$ & $-0.9$ &$0.4$ &$1.05 \pm 0.04$ &$1.03$ \\
$18 \div 34$ &$0.797$ &$11.6$ & $4.8$ & $10.5$ &$1.1$ & $2.1$ & $-5.1$ &$0.1$ & $0.6$ &$-0.1$ & $-2.9$ &$-2.5$ & $-1.3$ &$1.4$ & $-0.6$ &$0.2$ &$1.06 \pm 0.04$ &$1.03$ \\
$34 \div 100$ &$0.164$ &$12.3$ & $6.2$ & $10.6$ &$0.9$ & $2.3$ & $-5.0$ &$-0.2$ & $0.5$ &$-0.6$ & $-2.7$ &$-2.9$ & $-1.5$ &$1.6$ & $-0.8$ &$0.1$ &$1.06 \pm 0.04$ &$1.03$ \\
\hline
$y$ & $d\sigma/dy$ & $\delta_{tot}$ & $\delta_{stat}$ & $\delta_{sys}$ & $\delta_{\theta}$ & $\delta_{E}$ & $\delta_{HFS}$
        & $\delta_{\qsq}$ & $\delta_{\xpom}$ & $\delta_{\beta}$ & $\delta_{p*_{T,1}}$
        & $\delta_{\zpom}$ & $\delta_{x_{dijet}}$ & $\delta_{\Delta\eta*}$ & $\delta_{t}$
        & $\delta_{bgr}$ & $1+\delta_{had}$ & $1+\delta_{rad}$\\
& $\rm [pb]$ & $[\%]$ & $[\%]$ & $[\%]$ & $[\%]$ & $[\%]$ & $[\%]$ & $[\%]$ &  $[\%]$ &  $[\%]$ &  $[\%]$ &  $[\%]$ &
        $[\%]$ &  $[\%]$ &  $[\%]$ & $[\%]$ & & \\
\hline
$0.10 \div 0.22$ &$113$ &$18.4$ & $6.5$ & $17.2$ &$2.1$ & $0.2$ & $-8.7$ &$-3.6$ & $-0.2$ &$-4.2$ & $-3.5$ &$-8.9$ & $-3.6$ &$3.9$ & $-1.5$ &$0.6$ &$1.01 \pm 0.06$ &$1.07$ \\
$0.22 \div 0.34$ &$163$ &$12.7$ & $4.5$ & $11.9$ &$2.0$ & $1.1$ & $-5.9$ &$-2.0$ & $0.5$ &$-1.5$ & $-3.2$ &$-4.1$ & $-2.1$ &$1.4$ & $-0.9$ &$0.6$ &$1.02 \pm 0.04$ &$1.05$ \\
$0.34 \div 0.46$ &$144$ &$11.2$ & $4.3$ & $10.4$ &$1.6$ & $2.8$ & $-4.2$ &$-0.4$ & $0.8$ &$-0.1$ & $-3.1$ &$-2.3$ & $-1.3$ &$1.0$ & $-1.1$ &$0.3$ &$1.06 \pm 0.04$ &$1.04$ \\
$0.46 \div 0.58$ &$106$ &$11.2$ & $5.0$ & $10.0$ &$1.2$ & $3.2$ & $-3.2$ &$0.7$ & $0.8$ &$0.9$ & $-3.1$ &$-1.0$ & $-0.6$ &$1.9$ & $-0.3$ &$0.4$ &$1.13 \pm 0.03$ &$1.02$ \\
$0.58 \div 0.70$ &$76.5$ &$12.4$ & $7.0$ & $10.2$ &$0.7$ & $4.3$ & $-2.3$ &$1.0$ & $0.6$ &$1.6$ & $-3.3$ &$0.3$ & $0.4$ &$1.2$ & $-1.5$ &$0.2$ &$1.17 \pm 0.02$ &$0.97$ \\
\hline
$\xpom$ & $d\sigma/d\xpom$ & $\delta_{tot}$ & $\delta_{stat}$ & $\delta_{sys}$ & $\delta_{\theta}$ & $\delta_{E}$ & $\delta_{HFS}$
        & $\delta_{\qsq}$ & $\delta_{\xpom}$ & $\delta_{\beta}$ & $\delta_{p*_{T,1}}$
        & $\delta_{\zpom}$ & $\delta_{x_{dijet}}$ & $\delta_{\Delta\eta*}$ & $\delta_{t}$
        & $\delta_{bgr}$ & $1+\delta_{had}$ & $1+\delta_{rad}$\\
& $\rm [pb]$ & $[\%]$ & $[\%]$ & $[\%]$ & $[\%]$ & $[\%]$ & $[\%]$ & $[\%]$ &  $[\%]$ &  $[\%]$ &  $[\%]$ &  $[\%]$ &
        $[\%]$ &  $[\%]$ &  $[\%]$ & $[\%]$ & & \\
\hline
$-2.30 \div -2.10$ &$14.2$ &$42.0$ & $36.2$ & $21.1$ &$1.8$ & $3.9$ & $-9.3$ &$-2.7$ & $4.0$ &$-3.7$ & $-5.8$ &$-11.4$ & $-4.7$ &$7.6$ & $-1.2$ &$0.7$ &$1.17 \pm 0.13$ &$1.06$ \\
$-2.10 \div -1.90$ &$53.5$ &$14.7$ & $8.9$ & $11.7$ &$1.6$ & $2.4$ & $-5.6$ &$-0.6$ & $1.2$ &$-0.8$ & $-3.2$ &$-3.7$ & $-1.8$ &$2.3$ & $-1.4$ &$0.0$ &$1.10 \pm 0.08$ &$1.04$ \\
$-1.90 \div -1.70$ &$111$ &$11.6$ & $5.5$ & $10.2$ &$1.5$ & $1.3$ & $-4.5$ &$-1.1$ & $0.1$ &$-0.2$ & $-3.5$ &$-1.5$ & $-1.0$ &$1.4$ & $-1.1$ &$0.0$ &$1.06 \pm 0.04$ &$1.04$ \\
$-1.70 \div -1.52$ &$196$ &$10.9$ & $4.9$ & $9.8$ &$1.3$ & $2.5$ & $-3.6$ &$-1.0$ & $-0.5$ &$0.5$ & $-3.4$ &$-0.0$ & $-0.3$ &$-0.5$ & $-0.5$ &$0.8$ &$1.03 \pm 0.03$ &$1.03$ \\
\hline
$\zpom$ & $d\sigma/d\zpom$ & $\delta_{tot}$ & $\delta_{stat}$ & $\delta_{sys}$ & $\delta_{\theta}$ & $\delta_{E}$ & $\delta_{HFS}$
        & $\delta_{\qsq}$ & $\delta_{\xpom}$ & $\delta_{\beta}$ & $\delta_{p*_{T,1}}$
        & $\delta_{\zpom}$ & $\delta_{x_{dijet}}$ & $\delta_{\Delta\eta*}$ & $\delta_{t}$
        & $\delta_{bgr}$ & $1+\delta_{had}$ & $1+\delta_{rad}$\\
& $\rm [pb]$ & $[\%]$ & $[\%]$ & $[\%]$ & $[\%]$ & $[\%]$ & $[\%]$ & $[\%]$ &  $[\%]$ &  $[\%]$ &  $[\%]$ &  $[\%]$ &
        $[\%]$ &  $[\%]$ &  $[\%]$ & $[\%]$ & & \\
\hline
$0.00 \div 0.22$ &$70.4$ &$20.3$ & $9.3$ & $18.0$ &$1.4$ & $3.4$ & $-4.0$ &$1.2$ & $-0.5$ &$4.6$ & $-2.4$ &$12.0$ & $4.1$ &$6.5$ & $-0.5$ &$0.8$ &$1.10 \pm 0.03$ &$1.06$ \\
$0.22 \div 0.40$ &$132$ &$11.9$ & $6.3$ & $10.1$ &$1.5$ & $3.0$ & $-1.2$ &$-0.9$ & $0.3$ &$-0.1$ & $-3.9$ &$-2.3$ & $-2.2$ &$1.0$ & $-0.9$ &$0.4$ &$1.07 \pm 0.02$ &$1.04$ \\
$0.40 \div 0.60$ &$89.7$ &$14.9$ & $6.8$ & $13.3$ &$1.2$ & $1.6$ & $-9.1$ &$-1.3$ & $0.8$ &$-1.2$ & $-2.8$ &$-3.9$ & $-1.4$ &$0.5$ & $-0.6$ &$0.3$ &$1.10 \pm 0.03$ &$1.02$ \\
$0.60 \div 0.80$ &$54.8$ &$14.9$ & $7.5$ & $12.9$ &$2.5$ & $1.9$ & $7.6$ &$-1.4$ & $0.9$ &$-1.2$ & $-3.2$ &$-4.2$ & $-1.4$ &$0.2$ & $-2.0$ &$0.1$ &$1.10 \pm 0.10$ &$1.02$ \\
$0.80 \div 1.00$ &$19.9$ &$45.0$ & $11.4$ & $43.5$ &$0.8$ & $0.6$ & $-42.1$ &$-1.9$ & $1.3$ &$-2.4$ & $-2.5$ &$-5.1$ & $-2.0$ &$3.0$ & $-1.5$ &$0.6$ &$0.57 \pm 0.10$ &$1.00$ \\
\hline
\end{tabular}
\end{minipage}}
\caption{Diffractive DIS dijet cross section measured differentially as a function of $\qsq$, $y$, $\log\xpom$ and $\zpom$.
        The statistical $\delta_{stat}$ and systematic $\delta_{sys}$ uncertainties are given together with the total uncertainty $\delta_{tot}$. The next 12 columns represent $+1\sigma$ shifts for the systematic error contributions from:
        electron polar angle measurement $\delta_{\theta}$,
        electron energy scale $\delta_{E}$, HFS energy scale $\delta_{HFS}$,
        model uncertainties $\delta_{\qsq}$,  $\delta_{\xpom}$, $\delta_{\beta}$, $\delta_{p*_{T,1}}$,
	$\delta_{\zpom}$, $\delta_{x_{dijet}}$, $\delta_{\Delta\eta*}$ and $\delta_{t}$
        and the background normalisation uncertainty $\delta_{bgr}$.
	The global normalisation uncertainty of $7.8\%$ is not listed explicitly but is included in the total 
	systematic uncertainty $\delta_{sys}$.
        The 
        last two column show the correction factors for hadronisation and QED radiation, respectively.
        }
\label{tab:tab1}
\end{table}
\end{landscape}

\newpage
\clearpage
\begin{landscape}
\begin{table}
\resizebox{.8\textwidth}{!}{\begin{minipage}{\textwidth}
\begin{tabular}{|c|c|c|c|c||c|c|c|c|c|c|c|c|c|c|c|c||c||c|}
\hline
$p^{\ast}_{T,1}$ & $d\sigma/dp^{\ast}_{T,1}$ & $\delta_{tot}$ & $\delta_{stat}$ & $\delta_{sys}$ & $\delta_{\theta}$ & $\delta_{E}$ & $\delta_{HFS}$
        & $\delta_{\qsq}$ & $\delta_{\xpom}$ & $\delta_{\beta}$ & $\delta_{p*_{T,1}}$
        & $\delta_{\zpom}$ & $\delta_{x_{dijet}}$ & $\delta_{\Delta\eta*}$ & $\delta_{t}$
        & $\delta_{bgr}$ & $1+\delta_{had}$ & $1+\delta_{rad}$\\
$[GeV]$ & $\rm [pb/GeV]$ & $[\%]$ & $[\%]$ & $[\%]$ & $[\%]$ & $[\%]$ & $[\%]$ & $[\%]$ &  $[\%]$ &  $[\%]$ &  $[\%]$ &  $[\%]$ &
        $[\%]$ &  $[\%]$ &  $[\%]$ & $[\%]$ & & \\
\hline
$5.50 \div 7.00$ &$30.8$ &$9.6$ & $3.2$ & $9.0$ &$1.4$ & $1.4$ & $-3.1$ &$-0.9$ & $0.5$ &$-0.6$ & $-1.3$ &$-1.7$ & $-0.9$ &$0.6$ & $-0.8$ &$0.1$ &$1.05 \pm 0.05$ &$1.03$ \\
$7.00 \div 9.00$ &$10.5$ &$11.8$ & $6.1$ & $10.0$ &$1.3$ & $3.0$ & $-4.6$ &$-0.6$ & $0.8$ &$-0.4$ & $-1.4$ &$-1.5$ & $-0.9$ &$1.2$ & $-1.0$ &$0.7$ &$1.06 \pm 0.04$ &$1.04$ \\
$9.00 \div 15.00$ &$1.07$ &$19.6$ & $12.7$ & $14.9$ &$1.3$ & $2.3$ & $-9.8$ &$-0.1$ & $1.0$ &$-0.1$ & $-4.2$ &$-2.7$ & $-1.1$ &$5.3$ & $-1.5$ &$0.7$ &$1.04 \pm 0.03$ &$1.06$ \\
\hline
$p^{\ast}_{T,2}$ & $d\sigma/dp^{\ast}_{T,2}$ & $\delta_{tot}$ & $\delta_{stat}$ & $\delta_{sys}$ & $\delta_{\theta}$ & $\delta_{E}$ & $\delta_{HFS}$
        & $\delta_{\qsq}$ & $\delta_{\xpom}$ & $\delta_{\beta}$ & $\delta_{p*_{T,1}}$
        & $\delta_{\zpom}$ & $\delta_{x_{dijet}}$ & $\delta_{\Delta\eta*}$ & $\delta_{t}$
        & $\delta_{bgr}$ & $1+\delta_{had}$ & $1+\delta_{rad}$\\
$[GeV]$ & $\rm [pb/GeV]$ & $[\%]$ & $[\%]$ & $[\%]$ & $[\%]$ & $[\%]$ & $[\%]$ & $[\%]$ &  $[\%]$ &  $[\%]$ &  $[\%]$ &  $[\%]$ &
        $[\%]$ &  $[\%]$ &  $[\%]$ & $[\%]$ & & \\
\hline
$4.00 \div 6.50$ &$22.3$ &$10.4$ & $3.7$ & $9.7$ &$1.5$ & $2.3$ & $-3.8$ &$-1.0$ & $0.7$ &$-0.8$ & $-1.2$ &$-2.0$ & $-1.2$ &$1.2$ & $-0.8$ &$0.1$ &$1.10 \pm 0.06$ &$1.03$ \\
$6.50 \div 9.00$ &$5.67$ &$12.2$ & $6.9$ & $10.1$ &$1.2$ & $2.0$ & $-4.9$ &$-0.6$ & $0.6$ &$-0.2$ & $-2.6$ &$-1.3$ & $-0.5$ &$1.2$ & $-1.0$ &$0.6$ &$0.97 \pm 0.02$ &$1.04$ \\
$9.00 \div 15.00$ &$0.539$ &$18.2$ & $12.3$ & $13.4$ &$1.1$ & $1.5$ & $-7.8$ &$0.5$ & $0.6$ &$0.8$ & $-6.2$ &$-2.4$ & $-0.9$ &$2.8$ & $-1.3$ &$0.1$ &$0.97 \pm 0.02$ &$1.06$ \\
\hline
$\langle p^{\ast}_{T}\rangle$ & $d\sigma/d\langle p^{\ast}_{T}\rangle$ & $\delta_{tot}$ & $\delta_{stat}$ & $\delta_{sys}$ & $\delta_{\theta}$ & $\delta_{E}$ & $\delta_{HFS}$
        & $\delta_{\qsq}$ & $\delta_{\xpom}$ & $\delta_{\beta}$ & $\delta_{p*_{T,1}}$
        & $\delta_{\zpom}$ & $\delta_{x_{dijet}}$ & $\delta_{\Delta\eta*}$ & $\delta_{t}$
        & $\delta_{bgr}$ & $1+\delta_{had}$ & $1+\delta_{rad}$\\
$[GeV]$ & $\rm [pb/GeV]$ & $[\%]$ & $[\%]$ & $[\%]$ & $[\%]$ & $[\%]$ & $[\%]$ & $[\%]$ &  $[\%]$ &  $[\%]$ &  $[\%]$ &  $[\%]$ &
        $[\%]$ &  $[\%]$ &  $[\%]$ & $[\%]$ & & \\
\hline
$4.75 \div 6.50$ &$27.6$ &$9.9$ & $3.5$ & $9.3$ &$1.5$ & $2.0$ & $-3.3$ &$-1.1$ & $0.5$ &$-0.8$ & $-1.0$ &$-1.9$ & $-1.0$ &$0.8$ & $-0.8$ &$0.1$ &$1.09 \pm 0.06$ &$1.03$ \\
$6.50 \div 9.00$ &$8.52$ &$11.3$ & $5.2$ & $10.0$ &$1.4$ & $2.4$ & $-5.0$ &$-0.4$ & $0.8$ &$-0.1$ & $-1.7$ &$-0.8$ & $-0.4$ &$1.5$ & $-1.1$ &$0.5$ &$1.01 \pm 0.03$ &$1.04$ \\ 
$9.00 \div 15.00$ &$0.701$ &$19.7$ & $13.4$ & $14.4$ &$0.7$ & $1.2$ & $-9.2$ &$-0.3$ & $0.7$ &$-0.2$ & $-5.4$ &$-3.5$ & $-1.3$ &$3.8$ & $-0.9$ &$0.5$ &$1.01 \pm 0.03$ &$1.06$ \\
\hline
$\Delta\eta^{\ast}$ & $d\sigma/d\Delta\eta^{\ast}$ & $\delta_{tot}$ & $\delta_{stat}$ & $\delta_{sys}$ & $\delta_{\theta}$ & $\delta_{E}$ & $\delta_{HFS}$
        & $\delta_{\qsq}$ & $\delta_{\xpom}$ & $\delta_{\beta}$ & $\delta_{p*_{T,1}}$
        & $\delta_{\zpom}$ & $\delta_{x_{dijet}}$ & $\delta_{\Delta\eta*}$ & $\delta_{t}$
        & $\delta_{bgr}$ & $1+\delta_{had}$ & $1+\delta_{rad}$\\
& $\rm [pb]$ & $[\%]$ & $[\%]$ & $[\%]$ & $[\%]$ & $[\%]$ & $[\%]$ & $[\%]$ &  $[\%]$ &  $[\%]$ &  $[\%]$ &  $[\%]$ &
        $[\%]$ &  $[\%]$ &  $[\%]$ & $[\%]$ & & \\
\hline
$0.00 \div 0.15$ &$51.6$ &$17.9$ & $9.5$ & $15.1$ &$1.6$ & $2.8$ & $-4.4$ &$-1.0$ & $1.0$ &$-0.8$ & $-3.8$ &$-2.3$ & $-1.4$ &$10.6$ & $-1.2$ &$0.1$ &$1.04 \pm 0.03$ &$1.03$ \\
$0.15 \div 0.40$ &$57.8$ &$14.1$ & $7.3$ & $12.1$ &$1.2$ & $1.0$ & $-5.1$ &$-0.9$ & $0.9$ &$-0.7$ & $-3.0$ &$-2.1$ & $-1.5$ &$6.2$ & $-1.1$ &$0.2$ &$1.05 \pm 0.03$ &$1.04$ \\
$0.40 \div 0.80$ &$45.1$ &$12.5$ & $5.7$ & $11.1$ &$1.9$ & $2.5$ & $-4.3$ &$-0.9$ & $0.8$ &$-0.5$ & $-3.8$ &$-2.2$ & $-1.2$ &$3.2$ & $-1.3$ &$0.5$ &$1.06 \pm 0.04$ &$1.04$ \\
$0.80 \div 1.30$ &$33.9$ &$12.3$ & $5.5$ & $10.9$ &$1.7$ & $2.4$ & $-4.7$ &$-1.0$ & $0.5$ &$-0.3$ & $-3.7$ &$-2.4$ & $-1.0$ &$-2.5$ & $-0.6$ &$0.3$ &$1.07 \pm 0.05$ &$1.03$ \\
$1.30 \div 3.00$ &$9.29$ &$15.0$ & $6.7$ & $13.4$ &$1.2$ & $3.4$ & $-5.3$ &$-1.0$ & $0.2$ &$-0.0$ & $-2.8$ &$-3.4$ & $-1.2$ &$-7.4$ & $-1.1$ &$0.3$ &$1.04 \pm 0.06$ &$1.03$ \\
\hline
\end{tabular}
\end{minipage}}
\caption{Diffractive DIS dijet cross section measured differentially as a function of $p^{\ast}_{T,1}$, $p^{\ast}_{T,2}$, $\langle p^{\ast}_{T}\rangle$ 
	and $\Delta\eta^{\ast}$.
        The statistical $\delta_{stat}$ and systematic $\delta_{sys}$ uncertainties are given together with the total uncertainty $\delta_{tot}$.
	Further details are given in Table~\ref{tab:tab1}.
        }
\label{tab:tab2}
\end{table}
\end{landscape}

\newpage
\clearpage
\begin{landscape}
\begin{table}[!ht]
\resizebox{.8\textwidth}{!}{\begin{minipage}{\textwidth}
\begin{tabular}{|c|c|c|c|c|c||c|c|c|c|c|c|c|c|c|c|c|c||c||c|}
\hline
$\zpom$ & $\qsq$ & $\frac{d^{2}\sigma}{d\zpom d\qsq}$ & $\delta_{tot}$ & $\delta_{stat}$ & $\delta_{sys}$ & $\delta_{\theta}$ & $\delta_{E}$ & $\delta_{HFS}$
& $\delta_{\qsq}$ & $\delta_{\xpom}$ & $\delta_{\beta}$ & $\delta_{p*_{T,1}}$
& $\delta_{\zpom}$ & $\delta_{x_{dijet}}$ & $\delta_{\Delta\eta*}$ & $\delta_{t}$
& $\delta_{bgr}$ & $1+\delta_{had}$ & $1+\delta_{rad}$\\
& $\rm [GeV^{2}]$ & $\rm [pb/GeV^{2}]$ & $[\%]$ & $[\%]$ & $[\%]$ & $[\%]$ & $[\%]$ & $[\%]$ & $[\%]$ & $[\%]$ & $[\%]$
& $[\%]$ & $[\%]$ & $[\%]$ & $[\%]$ & $[\%]$ & $[\%]$ & & \\
\hline
$0.0 \div 0.3$ &$4 \div 10$ &$7.67$ &$14.8$ & $7.7$ & $12.7$ &$2.0$ & $3.5$ & $-1.0$ &$1.6$ & $-0.3$ &$3.4$ & $-3.3$ &$6.5$ & $2.2$ &$3.1$ & $-1.3$ &$0.5$ &$1.08 \pm 0.03$ &$1.05$ \\
 & $10 \div 20$ &$2.40$ &$15.6$ & $10.0$ & $12.1$ &$0.7$ & $2.9$ & $-2.1$ &$-2.0$ & $-0.2$ &$1.1$ & $-3.2$ &$5.7$ & $1.5$ &$4.6$ & $-0.4$ &$0.7$ &$1.08 \pm 0.02$ &$1.05$ \\
 & $20 \div 40$ &$0.544$ &$27.6$ & $20.8$ & $18.2$ &$2.2$ & $4.7$ & $-3.3$ &$0.9$ & $-0.2$ &$4.7$ & $-3.5$ &$11.6$ & $2.8$ &$7.3$ & $-0.4$ &$0.2$ &$1.09 \pm 0.02$ &$1.05$ \\
 & $40 \div 100$ &$0.0994$ &$41.6$ & $35.7$ & $21.3$ &$1.2$ & $7.3$ & $-3.3$ &$-0.4$ & $-0.1$ &$4.1$ & $-4.1$ &$12.9$ & $3.1$ &$10.0$ & $-3.7$ &$2.0$ &$1.09 \pm 0.02$ &$1.06$ \\
$0.3 \div 0.5$ &$4 \div 10$ &$8.80$ &$18.4$ & $9.3$ & $15.9$ &$2.0$ & $3.1$ & $-8.6$ &$-0.7$ & $1.2$ &$-1.0$ & $-5.3$ &$-7.4$ & $-3.9$ &$-0.6$ & $-1.5$ &$0.1$ &$1.08 \pm 0.02$ &$1.03$ \\
 & $10 \div 20$ &$2.31$ &$19.9$ & $13.6$ & $14.5$ &$2.0$ & $2.4$ & $-6.9$ &$-0.4$ & $0.7$ &$-1.5$ & $-3.3$ &$-7.7$ & $-4.1$ &$-0.0$ & $-0.4$ &$1.0$ &$1.08 \pm 0.02$ &$1.03$ \\
 & $20 \div 40$ &$1.12$ &$17.0$ & $12.6$ & $11.4$ &$-0.4$ & $3.2$ & $-3.7$ &$0.2$ & $0.5$ &$-0.7$ & $-3.0$ &$-5.3$ & $-2.8$ &$-0.3$ & $-0.0$ &$0.2$ &$1.08 \pm 0.03$ &$1.02$ \\
 & $40 \div 100$ &$0.264$ &$20.1$ & $17.1$ & $10.6$ &$0.7$ & $2.6$ & $-4.4$ &$-0.2$ & $0.4$ &$-0.9$ & $-2.1$ &$-3.4$ & $-2.1$ &$1.4$ & $1.3$ &$0.2$ &$1.07 \pm 0.03$ &$1.03$ \\
$0.5 \div 0.7$ &$4 \div 10$ &$4.50$ &$17.8$ & $13.3$ & $11.8$ &$3.3$ & $1.9$ & $6.5$ &$-1.4$ & $1.1$ &$-0.7$ & $-3.1$ &$-2.8$ & $-0.4$ &$0.9$ & $-0.7$ &$0.2$ &$1.14 \pm 0.06$ &$1.03$ \\
 & $10 \div 20$ &$1.86$ &$15.2$ & $11.8$ & $9.6$ &$0.8$ & $1.0$ & $3.1$ &$-0.5$ & $0.5$ &$-0.4$ & $-3.1$ &$-2.6$ & $-0.6$ &$0.2$ & $-1.8$ &$0.1$ &$1.12 \pm 0.06$ &$1.02$ \\
 & $20 \div 40$ &$0.703$ &$16.2$ & $13.5$ & $8.9$ &$2.0$ & $0.8$ & $-0.2$ &$-0.7$ & $0.5$ &$-0.6$ & $-2.2$ &$-2.3$ & $-0.7$ &$0.2$ & $-1.6$ &$0.2$ &$1.12 \pm 0.06$ &$1.02$ \\
 & $40 \div 100$ &$0.109$ &$31.9$ & $29.7$ & $11.4$ &$2.2$ & $-0.8$ & $3.2$ &$-0.8$ & $0.1$ &$-1.3$ & $-1.4$ &$-4.0$ & $-1.0$ &$-1.6$ & $-5.6$ &$0.1$ &$1.12 \pm 0.06$ &$1.01$ \\
$0.7 \div 1.0$ &$4 \div 10$ &$1.99$ &$27.8$ & $11.7$ & $25.2$ &$2.2$ & $2.9$ & $-21.9$ &$-1.6$ & $1.8$ &$-1.9$ & $-3.8$ &$-6.5$ & $-2.6$ &$1.7$ & $-2.9$ &$0.3$ &$0.79 \pm 0.11$ &$1.02$ \\
 & $10 \div 20$ &$0.639$ &$26.9$ & $11.2$ & $24.5$ &$1.4$ & $0.4$ & $-22.1$ &$-0.4$ & $1.1$ &$-1.6$ & $-1.7$ &$-5.2$ & $-2.0$ &$2.4$ & $-1.8$ &$0.3$ &$0.81 \pm 0.11$ &$1.01$ \\
 & $20 \div 40$ &$0.248$ &$22.4$ & $13.0$ & $18.2$ &$0.9$ & $1.6$ & $-15.3$ &$-0.3$ & $0.9$ &$-1.1$ & $-2.6$ &$-4.3$ & $-1.6$ &$1.1$ & $-0.3$ &$0.0$ &$0.85 \pm 0.11$ &$1.00$ \\
 & $40 \div 100$ &$0.0968$ &$18.5$ & $13.3$ & $13.0$ &$0.3$ & $2.1$ & $-9.0$ &$-0.5$ & $0.4$ &$-1.1$ & $-2.2$ &$-3.4$ & $-1.5$ &$1.3$ & $0.4$ &$0.5$ &$0.89 \pm 0.10$ &$1.01$ \\
\hline
\end{tabular}
\end{minipage}}
\caption{Diffractive DIS dijet cross section measured differentially
        as a function of $\zpom$ and $\qsq$.
        The statistical $\delta_{stat}$ and systematic $\delta_{sys}$ uncertainties are given together with the total uncertainty $\delta_{tot}$.
        Further details are given in Table~\ref{tab:tab1}.
	}
\label{tab:tab3}
\end{table}
\end{landscape}

\newpage
\clearpage
\begin{landscape}
\begin{table}[!ht]
\resizebox{.8\textwidth}{!}{\begin{minipage}{\textwidth}
\begin{tabular}{|c|c|c|c|c|c||c|c|c|c|c|c|c|c|c|c|c|c||c||c|}
\hline
$p^{\ast}_{\rm T,1}$ & $\qsq$ & $\frac{d^{2}\sigma}{dp^{\ast}_{\rm T,1}d\qsq}$  & $\delta_{tot}$ & $\delta_{stat}$ & $\delta_{sys}$
& $\delta_{\theta}$ & $\delta_{E}$ & $\delta_{HFS}$
& $\delta_{\qsq}$ & $\delta_{\xpom}$ & $\delta_{\beta}$ & $\delta_{p*_{T,1}}$
& $\delta_{\zpom}$ & $\delta_{x_{dijet}}$ & $\delta_{\Delta\eta*}$ & $\delta_{t}$
& $\delta_{bgr}$ & $1+\delta_{had}$ & $1+\delta_{rad}$\\
$\rm [GeV^{2}]$ & $\rm [GeV]$ & $\rm [pb/GeV^{3}]$ & $[\%]$ & $[\%]$ & $[\%]$ & $[\%]$ & $[\%]$ & $[\%]$ & $[\%]$ & $[\%]$ & $[\%]$
& $[\%]$ & $[\%]$ & $[\%]$ & $[\%]$ & $[\%]$ & $[\%]$ & & \\
\hline
$5.5 \div 7.0$ &$4 \div 6$ &$3.35$ &$15.6$ & $9.1$ & $12.7$ &$0.4$ & $5.9$ & $-1.2$ &$2.7$ & $1.1$ &$3.4$ & $-1.3$ &$5.8$ & $2.9$ &$0.6$ & $-1.0$ &$0.1$ &$1.05 \pm 0.05$ &$1.04$ \\
 & $6 \div 10$ &$1.84$ &$12.7$ & $7.1$ & $10.5$ &$3.0$ & $0.8$ & $-4.7$ &$-1.5$ & $0.3$ &$-1.0$ & $-1.8$ &$-2.9$ & $-1.3$ &$0.7$ & $-1.6$ &$0.1$ &$1.05 \pm 0.05$ &$1.02$ \\
 & $10 \div 18$ &$0.834$ &$12.3$ & $7.2$ & $9.9$ &$1.2$ & $0.9$ & $-3.2$ &$-1.1$ & $0.5$ &$-1.4$ & $-1.6$ &$-3.9$ & $-1.7$ &$0.6$ & $-1.1$ &$0.2$ &$1.05 \pm 0.05$ &$1.02$ \\
 & $18 \div 34$ &$0.344$ &$13.3$ & $8.6$ & $10.1$ &$1.5$ & $0.4$ & $-5.1$ &$0.3$ & $0.5$ &$-0.4$ & $-1.4$ &$-2.6$ & $-1.7$ &$1.1$ & $-0.0$ &$0.0$ &$1.06 \pm 0.05$ &$1.03$ \\
 & $34 \div 100$ &$0.0613$ &$15.8$ & $11.7$ & $10.6$ &$1.5$ & $0.7$ & $-5.0$ &$-0.4$ & $0.5$ &$-1.1$ & $-1.7$ &$-3.4$ & $-2.0$ &$0.9$ & $-1.6$ &$0.1$ &$1.07 \pm 0.04$ &$1.02$ \\
$7.0 \div 9.0$ &$4 \div 6$ &$1.23$ &$18.6$ & $15.1$ & $10.9$ &$-0.1$ & $3.0$ & $-6.1$ &$1.3$ & $1.2$ &$1.5$ & $-2.5$ &$0.1$ & $-0.7$ &$0.0$ & $-0.3$ &$0.5$ &$1.06 \pm 0.04$ &$1.05$ \\
 & $6 \div 10$ &$0.578$ &$16.4$ & $12.9$ & $10.1$ &$1.9$ & $3.3$ & $-3.9$ &$1.1$ & $0.8$ &$0.0$ & $-0.6$ &$-2.1$ & $-1.3$ &$1.5$ & $-0.8$ &$0.4$ &$1.06 \pm 0.04$ &$1.05$ \\
 & $10 \div 18$ &$0.287$ &$16.6$ & $12.6$ & $10.7$ &$0.4$ & $3.1$ & $-6.0$ &$-0.3$ & $0.7$ &$-0.1$ & $-1.3$ &$-0.3$ & $-0.4$ &$2.3$ & $0.7$ &$0.6$ &$1.06 \pm 0.05$ &$1.04$ \\
 & $18 \div 34$ &$0.100$ &$20.4$ & $17.6$ & $10.3$ &$0.3$ & $5.2$ & $-3.5$ &$-0.2$ & $0.8$ &$-0.2$ & $-0.5$ &$-1.0$ & $-0.4$ &$0.0$ & $-1.5$ &$0.7$ &$1.07 \pm 0.04$ &$1.04$ \\
 & $34 \div 100$ &$0.0276$ &$19.9$ & $17.4$ & $9.6$ &$-0.6$ & $3.7$ & $-3.1$ &$-0.6$ & $0.5$ &$-1.0$ & $-0.6$ &$-1.6$ & $-0.4$ &$1.0$ & $1.5$ &$0.6$ &$1.06 \pm 0.06$ &$1.04$ \\
$9.0 \div 15.0$ &$4 \div 6$ &$0.122$ &$30.1$ & $26.6$ & $14.2$ &$7.8$ & $0.5$ & $-5.5$ &$-0.1$ & $0.7$ &$0.3$ & $-5.5$ &$-1.2$ & $-0.5$ &$3.3$ & $-2.2$ &$0.8$ &$1.04 \pm 0.03$ &$1.06$ \\
 & $6 \div 10$ &$0.0511$ &$30.4$ & $24.7$ & $17.8$ &$1.9$ & $1.1$ & $-12.4$ &$-0.6$ & $1.3$ &$-0.0$ & $-6.3$ &$-2.8$ & $-0.9$ &$6.1$ & $-3.0$ &$0.4$ &$1.03 \pm 0.03$ &$1.05$ \\
 & $10 \div 18$ &$0.0207$ &$35.5$ & $30.0$ & $19.0$ &$1.4$ & $1.7$ & $-11.6$ &$-1.1$ & $1.0$ &$-1.4$ & $-6.5$ &$-6.4$ & $-2.9$ &$5.6$ & $-6.0$ &$0.5$ &$1.03 \pm 0.02$ &$1.05$ \\
 & $18 \div 34$ &$0.0160$ &$24.5$ & $20.1$ & $14.0$ &$1.6$ & $1.8$ & $-8.2$ &$0.0$ & $0.6$ &$-0.1$ & $-4.6$ &$-2.6$ & $-0.6$ &$5.2$ & $-2.5$ &$0.4$ &$1.04 \pm 0.04$ &$1.06$ \\
 & $34 \div 100$ &$0.0034$ &$31.9$ & $27.8$ & $15.7$ &$3.6$ & $0.3$ & $-9.0$ &$1.8$ & $0.7$ &$1.0$ & $-3.5$ &$-3.1$ & $-1.5$ &$6.0$ & $-4.9$ &$1.0$ &$1.05 \pm 0.03$ &$1.07$ \\
\hline
\end{tabular}
\end{minipage}}
\caption{Diffractive DIS dijet cross section measured differentially
        as a function of $p^{\ast}_{\rm T,1}$ and $\qsq$.
        The statistical $\delta_{stat}$ and systematic $\delta_{sys}$ uncertainties are given together with the total uncertainty $\delta_{tot.}$.
        Further details are given in Table~\ref{tab:tab1}.
	}
\label{tab:tab5}
\end{table}
\end{landscape}

\clearpage
\begin{table}[!ht]
\resizebox{.8\textwidth}{!}{\begin{minipage}{\textwidth}
\centering
\begin{tabular}{|c|c|ccccc|}
\hline
$\qsq\ \rm [GeV]$ & $\#Bin$ & $1$ & $2$ & $3$ & $4$ & $5$\\
\hline
$4 \div 6$ & $1$ &$100$ & $-5$ & $5$ &  & \\
$6 \div 10$ & $2$ & & $100$ & $1$ & $1$ & \\
$10 \div 18$ & $3$ & &  & $100$ & $-2$ & $1$\\
$18 \div 34$ & $4$ & &  &  & $100$ & $8$\\
$34 \div 100$ & $5$ & &  &  &  & $100$\\
\hline
$y$ & $\#Bin$ & $1$ & $2$ & $3$ & $4$ & $5$\\
\hline
$0.1 \div 0.2$ & $1$ &$100$ & $-7$ & $8$ & $5$ & $4$\\
$0.2 \div 0.3$ & $2$ & & $100$ & $-6$ & $8$ & $4$\\
$0.3 \div 0.5$ & $3$ & &  & $100$ & $-4$ & $7$\\
$0.5 \div 0.6$ & $4$ & &  &  & $100$ & $-10$\\
$0.6 \div 0.7$ & $5$ & &  &  &  & $100$\\
\hline
$\xpom$ & $\#Bin$ & $1$ & $2$ & $3$ & $4$ &\\
\hline
$-2.30 \div -2.10$ & $1$ &$100$ & $-55$ & $17$ & $-2$ &\\
$-2.10 \div -1.90$ & $2$ & & $100$ & $-41$ & $11$ &\\
$-1.90 \div -1.70$ & $3$ & &  & $100$ & $-31$ &\\
$-1.70 \div -1.52$ & $4$ & &  &  & $100$ &\\
\hline
$\zpom$ & $\#Bin$ & $1$ & $2$ & $3$ & $4$ & $5$\\
\hline
$0.0 \div 0.2$ & $1$ &$100$ & $-24$ & $8$ & $1$ & \\
$0.2 \div 0.4$ & $2$ & & $100$ & $-31$ & $10$ & $-2$\\
$0.4 \div 0.6$ & $3$ & &  & $100$ & $-45$ & $17$\\
$0.6 \div 0.8$ & $4$ & &  &  & $100$ & $-52$\\
$0.8 \div 1.0$ & $5$ & &  &  &  & $100$\\
\hline
\end{tabular}
\end{minipage}}
\caption{Correlation coefficients between data points for the single-differential measurements in 
	$\qsq$, $y$, $\xpom$ and $\zpom$.
	The values are given in per cent.}
\label{tab:tabrho1}
\end{table}

\clearpage
\begin{table}[!ht]
\resizebox{.8\textwidth}{!}{\begin{minipage}{\textwidth}
\centering
\begin{tabular}{|c|c|ccccc|}
\hline
$p^{\ast}_{\rm T,1}\ \rm [GeV]$ & $\#Bin$
 & $1$ & $2$ & $3$ & &\\
\hline
$5.5 \div 7.0$ & $1$ &$100$ & $-26$ & $1$ &&\\
$7.0 \div 9.0$ & $2$ & & $100$ & $-54$ &&\\
$9.0 \div 15.0$ & $3$ & &  & $100$ &&\\
\hline
$p^{\ast}_{\rm T,2}\ \rm [GeV]$ & $\#Bin$ 
& $1$ & $2$ & $3$ & & \\
\hline
$4.0 \div 6.5$ & $1$ &$100$ & $-36$ & $13$ &&\\
$6.5 \div 9.0$ & $2$ & & $100$ & $-46$ &&\\
$9.0 \div 15.0$ & $3$ & &  & $100$ &&\\
\hline
$\langle p^{\ast}_{\rm T}\rangle\ \rm [GeV]$ & $\#Bin$
        & $1$ & $2$ & $3$ & &\\
\hline
$4.75 \div 6.50$ & $1$ &$100$ & $-33$ & $12$ &&\\
$6.50 \div 9.00$ & $2$ & & $100$ & $-49$ &&\\
$9.00 \div 15.00$ & $3$ & &  & $100$ &&\\
\hline
$\Delta\eta^{\ast}$ & $\#Bin$ 
& $1$ & $2$ & $3$ & $4$ & $5$ \\
\hline
$0.00 \div 0.15$ & $1$ &$100$ & $-49$ & $13$ & $1$ & $2$\\
$0.15 \div 0.40$ & $2$ & & $100$ & $-29$ & $9$ & $1$\\
$0.40 \div 0.80$ & $3$ & &  & $100$ & $-19$ & $7$\\
$0.80 \div 1.30$ & $4$ & &  &  & $100$ & $-20$\\
$1.30 \div 3.00$ & $5$ & &  &  &  & $100$\\
\hline
\end{tabular}
\end{minipage}}
\caption{Correlation coefficients between data points for the single-differential measurements in $p^{\ast}_{\rm T,1}$, $p^{\ast}_{\rm T,2}$, 
		  $\langle p^{\ast}_{\rm T}\rangle$ and $\Delta\eta^{\ast}$.
        The values are given in per cent.}
\label{tab:tabrho2}
\end{table}

\clearpage
\begin{table}[!ht]
\resizebox{\columnwidth}{!}{
\centering
\begin{tabular}{|c|c|c|cccccccccccccccc|}
\hline
$\zpom$ & $\qsq\ \rm [GeV^{2}]$ & $\#Bin$
 & $1$ & $2$ & $3$ & $4$ & $5$ & $6$ & $7$ & $8$ & $9$ & $10$ & $11$ & $12$ & $13$ & $14$ & $15$ & $16$\\
\hline
$0.0 \div 0.3$ & $4 \div 10$ & $1$ &$100$ & $1$ &  &  & $-32$ & $3$ &  &  & $11$ &  &  &  & $-3$ &  &  & \\
 & $10 \div 20$ & $2$ & & $100$ & $-2$ & $1$ & $3$ & $-41$ & $2$ &  &  & $19$ &  &  &  & $-5$ &  & \\
 & $20 \div 40$ & $3$ & &  & $100$ & $4$ &  & $2$ & $-37$ & $4$ &  &  & $20$ & $1$ &  &  & $-4$ & $1$\\
 & $40 \div 100$ & $4$ & &  &  & $100$ &  &  & $3$ & $-37$ &  &  & $1$ & $22$ &  &  & $1$ & $-5$\\
$0.3 \div 0.5$ & $4 \div 10$ & $5$ & &  &  &  & $100$ & $-3$ &  &  & $-46$ & $2$ &  &  & $15$ &  &  & \\
 & $10 \div 20$ & $6$ & &  &  &  &  & $100$ & $-3$ &  & $3$ & $-53$ & $2$ &  &  & $19$ &  & \\
 & $20 \div 40$ & $7$ & &  &  &  &  &  & $100$ & $-3$ &  & $2$ & $-51$ & $2$ &  & $-1$ & $17$ & \\
 & $40 \div 100$ & $8$ & &  &  &  &  &  &  & $100$ &  &  & $2$ & $-51$ &  &  &  & $21$\\
$0.5 \div 0.7$ & $4 \div 10$ & $9$ & &  &  &  &  &  &  &  & $100$ & $-5$ &  &  & $-47$ & $2$ &  & \\
 & $10 \div 20$ & $10$ & &  &  &  &  &  &  &  &  & $100$ & $-3$ &  & $1$ & $-46$ & $1$ & \\
 & $20 \div 40$ & $11$ & &  &  &  &  &  &  &  &  &  & $100$ & $-2$ &  & $2$ & $-44$ & $1$\\
 & $40 \div 100$ & $12$ & &  &  &  &  &  &  &  &  &  &  & $100$ &  &  & $1$ & $-51$\\
$0.7 \div 1.0$ & $4 \div 10$ & $13$ & &  &  &  &  &  &  &  &  &  &  &  & $100$ & $-4$ &  & \\
 & $10 \div 20$ & $14$ & &  &  &  &  &  &  &  &  &  &  &  &  & $100$ & $-5$ & \\
 & $20 \div 40$ & $15$ & &  &  &  &  &  &  &  &  &  &  &  &  &  & $100$ & $-2$\\
 & $40 \div 100$ & $16$ & &  &  &  &  &  &  &  &  &  &  &  &  &  &  & $100$\\
\hline
\end{tabular}
}
\caption{Correlation coefficients between data points for the double-differential measurement in $\zpom$ and $\qsq$.
        The values are given in per cent.}
\label{tab:tabrho3}
\end{table}

\begin{table}[!ht]
\resizebox{\columnwidth}{!}{
\begin{tabular}{|c|c|c|ccccccccccccccc|}
\hline
$p^{\ast}_{\rm T,1}\ \rm [GeV]$ & $\qsq\ \rm [GeV^{2}]$ & $\#Bin$
 & $1$ & $2$ & $3$ & $4$ & $5$ & $6$ & $7$ & $8$ & $9$ & $10$ & $11$ & $12$ & $13$ & $14$ & $15$\\
\hline
$5.5 \div 7.0$ & $4 \div 6$ & $1$ &$100$ & $-7$ & $1$ &  &  & $-44$ & $2$ &  &  &  & $13$ & $1$ & $1$ & $1$ & \\
 & $6 \div 10$ & $2$ & & $100$ & $-3$ &  &  & $3$ & $-57$ & $3$ & $-1$ &  &  & $17$ & $1$ & $1$ & $1$\\
 & $10 \div 18$ & $3$ & &  & $100$ & $-2$ & $1$ & $1$ & $3$ & $-59$ & $1$ & $-1$ & $2$ & $1$ & $22$ & $1$ & $1$\\
 & $18 \div 34$ & $4$ & &  &  & $100$ & $3$ & $-1$ &  &  & $-58$ & $1$ & $2$ & $1$ & $2$ & $25$ & $2$\\
 & $34 \div 100$ & $5$ & &  &  &  & $100$ &  &  &  & $1$ & $-56$ &  &  & $1$ & $2$ & $27$\\
$7.0 \div 9.0$ & $4 \div 6$ & $6$ & &  &  &  &  & $100$ & $-7$ & $3$ & $3$ & $1$ & $-60$ & $2$ & $-5$ & $-6$ & $-3$\\
 & $6 \div 10$ & $7$ & &  &  &  &  &  & $100$ & $-4$ & $2$ & $1$ & $3$ & $-57$ &  & $-3$ & $-2$\\
 & $10 \div 18$ & $8$ & &  &  &  &  &  &  & $100$ &  & $2$ & $-6$ &  & $-60$ & $-4$ & $-4$\\
 & $18 \div 34$ & $9$ & &  &  &  &  &  &  &  & $100$ & $1$ & $-7$ & $-3$ & $-6$ & $-62$ & $-3$\\
 & $34 \div 100$ & $10$ & &  &  &  &  &  &  &  &  & $100$ & $-4$ & $-2$ & $-5$ & $-5$ & $-64$\\
$9.0 \div 15.0$ & $4 \div 6$ & $11$ & &  &  &  &  &  &  &  &  &  & $100$ & $-5$ & $13$ & $14$ & $7$\\
 & $6 \div 10$ & $12$ & &  &  &  &  &  &  &  &  &  &  & $100$ &  & $6$ & $3$\\
 & $10 \div 18$ & $13$ & &  &  &  &  &  &  &  &  &  &  &  & $100$ & $10$ & $9$\\
 & $18 \div 34$ & $14$ & &  &  &  &  &  &  &  &  &  &  &  &  & $100$ & $8$\\
 & $34 \div 100$ & $15$ & &  &  &  &  &  &  &  &  &  &  &  &  &  & $100$\\
\hline
\end{tabular}
}
\caption{Correlation coefficients between data points for the double-differential measurement in $p^{\ast}_{\rm T,1}$ and $\qsq$.
        The values are given in per cent.}
\label{tab:tabrho5}
\end{table}

\newpage

\newpage

\newpage

\newpage
\clearpage

\newpage
\clearpage

\newpage
\clearpage

\newpage

\newpage
\clearpage

\newpage
\clearpage


\end{document}